\documentclass[a4paper,12pt]{article}
\usepackage{amsmath,amssymb,amsthm,graphicx}
\usepackage{enumitem}
\usepackage{color}
\usepackage{epsfig}
\usepackage{graphics}
\usepackage[font=small]{caption}
\usepackage[hang,flushmargin]{footmisc} 
\usepackage{float}
\usepackage{ifthen}
\usepackage[mathscr]{euscript}
\usepackage{natbib}
\usepackage{setspace}

\usepackage[left=2.24cm,right=2.24cm,bottom=2.5cm,top=2.5cm]{geometry}

\newcommand{\heading}[2]
{  \setcounter{page}{1}
   \begin{center}

   {\LARGE \textbf{#1}}
   \vspace{0.3cm}

   {\LARGE \textbf{#2}}
   \end{center}
   \vspace{3pt}
}

\newcommand{\authors}[4]
{  \parindent0pt
   \begin{center}
      \begin{minipage}[c][2cm][c]{5.25cm}
      \begin{center} 
      {\large #1} 
      \vspace{0.1cm}
      
      #2 
      \end{center}
      \end{minipage}
      \begin{minipage}[c][1.22cm][t]{5.25cm}
      \begin{center} 
      \vspace{1.8pt}
 
      {\large #3}
      \vspace{0.1cm}
      
      #4 
      \end{center}
      \end{minipage}
   \end{center}
}

\newcommand{\version}[1]
{  \begin{center}
   {\large #1}
   \end{center}
   \vspace{3pt}
}

\newcounter{zaehler}
\newlength{\labellength}
\newlength{\leftmarginlength}
\newlength{\labelseplength}

\newenvironment{listing}[2]
{   \setcounter{zaehler}{1}
    \settowidth{\labellength}{(#1\arabic{zaehler}#2)}
    \settowidth{\leftmarginlength}{(#1\arabic{zaehler}#2)-.}
    \settowidth{\labelseplength}{-.}
    \begin{list}{(#1\arabic{zaehler}#2)}
    {\usecounter{zaehler}
     \setlength{\labelwidth}{\labellength}
     \setlength{\labelsep}{\labelseplength}
     \setlength{\leftmargin}{\leftmarginlength} 
     \setlength{\listparindent}{0cm} 
     \setlength{\itemindent}{0cm} 
    }
}
{   \end{list}}

\newcommand{\Dpt}{D}
\newcommand{\Dsup}{\mathscr{D}}
\newcommand{\Dmax}{\mathbb{D}}
\newcommand{\Hpt}{H} 
\newcommand{\Hsup}{\mathscr{H}} 
\newcommand{\Hmax}{\mathbb{H}}

\newcommand{\reals}{\mathbb{R}}
\newcommand{\integers}{\mathbb{Z}}
\newcommand{\naturals}{\mathbb{N}}

\newcommand{\pr}{\mathbb{P}}        
\newcommand{\ex}{\mathbb{E}}        
\newcommand{\var}{\textnormal{Var}} 
\newcommand{\cov}{\textnormal{Cov}} 

\newcommand{\normal}{N}        

\newcommand{\dist}{d_{\mathcal{F}}} 

\newcommand{\convd}{\stackrel{d}{\longrightarrow}}              
\newcommand{\convp}{\stackrel{P}{\longrightarrow}}              
\newcommand{\convw}{\rightsquigarrow}                           

\theoremstyle{plain}

\newtheorem{theorem}{Theorem}[section]

\newtheorem{propositionA}{Proposition}[section]
\newtheorem{corollary}[theorem]{Corollary}

\newtheorem{lemmaA}{Lemma}[section]
\newtheorem{definition}{Definition}[section]

\newtheorem{example}{Example}

\allowdisplaybreaks[2]

\begin{document}

\heading{Detecting Gradual Changes}{in Locally Stationary Processes}
\renewcommand{\thefootnote}{*}
\authors{Michael Vogt}{University of Konstanz}{Holger Dette}{Ruhr-Universit\"at Bochum}
\version{\today}

\renewcommand{\thefootnote}{\arabic{footnote}}
\setcounter{footnote}{0}

\begin{abstract}
\noindent
In a wide range of applications, the stochastic properties of the observed time series change over time. The changes often occur gradually rather than abruptly: the properties are (approximately) constant for some time and then slowly start to change. In such situations, it is frequently of interest to locate the time point where the properties start to vary. In contrast to the analysis of abrupt changes, methods for detecting smooth or gradual change points are less developed and often require strong parametric assumptions. In this paper, we develop a fully nonparametric method to estimate a smooth change point in a locally stationary framework. We set up a general procedure which allows to deal with a wide variety of stochastic properties including the mean, (auto)covariances and higher-order moments. The theoretical part of the paper establishes the convergence rate of the new estimator. In addition, we examine its finite sample performance by means of a simulation study and illustrate the methodology by applications to temperature and financial return data. 
\end{abstract}
\vspace{5pt}

\vspace{3pt}


\renewcommand{\baselinestretch}{1.15}\normalsize

\section{Introduction}
\def\theequation{1.\arabic{equation}}
\setcounter{equation}{0}

In many applications, the stochastic properties of the observed time series such as the mean, the variance or the distribution change over time. In the classical structural break setting, the changes are abrupt: the stochastic properties are constant for some time and then suddenly jump to another value. In a number of situations, however, the changes occur gradually rather than abruptly: the properties are (approximately) constant for a while and then reach a time point where they slowly start to change. We refer to this time point as a smooth or gradual change point in what follows.

\begin{figure}[H]
\centering
\includegraphics[width=0.4\textwidth]{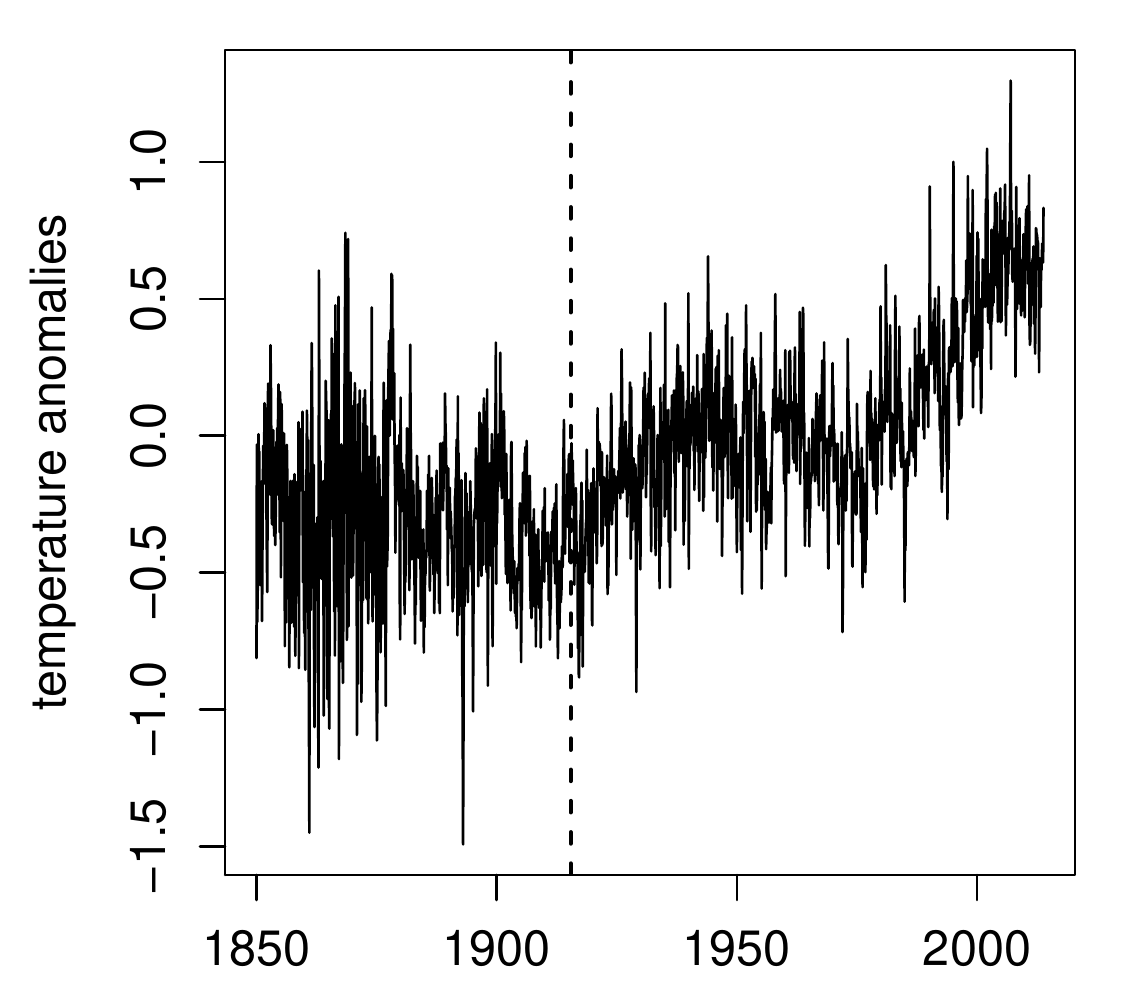}\hspace{0.5cm}
\includegraphics[width=0.4\textwidth]{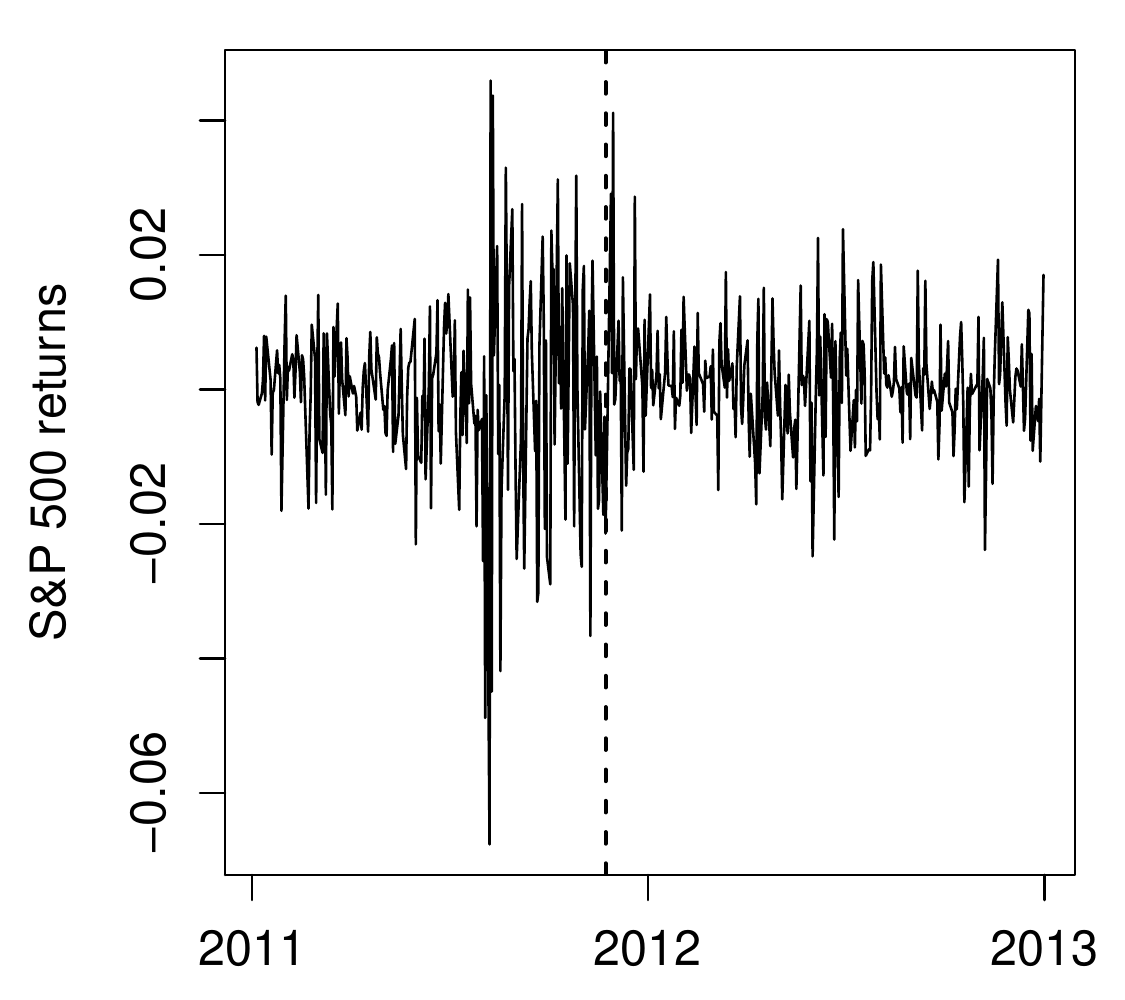}
\caption{The left-hand panel shows the monthly temperature anomalies of the northern hemisphere from 1850 to 2013 measured in $^\circ\rm{C}$. The right-hand panel depicts the daily returns of the S\&P 500 index from the beginning of 2011 to the end of 2012. The vertical lines in the figures indicate the gradual change points estimated by the method developed in this paper.}\label{fig1}
\end{figure}

Locating a smooth change point is important in a wide range of applications. As a first example, consider the monthly temperature anomalies (temperature deviations from a reference value) of the northern hemisphere from 1850 to 2013 which are displayed in the left-hand panel of Figure \ref{fig1}. Global mean temperature records over the last 150 years suggest that there has been a significant upward trend in the temperature [see \cite{Bloomfield1992} and \cite{Hansen2002} among others]. This upward trend which is commonly termed ``global warming'' is also visible in the time series of Figure \ref{fig1}. Inspecting the plot more closely, the mean temperature appears to be fairly constant at the beginning of the sample and then starts to gradually increase. An important issue is to detect the advent of ``global warming'', that is, the time point where the mean of the time series starts to trend upwards.

A second example can be found in the right-hand panel of Figure \ref{fig1} which shows the daily returns of the S\&P 500 stock index from the beginning of 2011 to the end of 2012. Inspecting the data, it is apparent that the volatility level changes over time. Moreover, the plot suggests that the volatility is roughly constant in 2012 but gradually increases before that. Denoting the present time point by $T$, practitioners are often interested in identifying the time interval $[t_0,T]$ where the volatility level is more or less constant. Put differently, they are interested in localizing the time point $t_0$ prior to which the volatility starts to substantially vary over time. Once the time point $t_0$ has been identified, it is common practice in volatility forecasting to fit a model to the data in the time span $[t_0,T]$ [see e.g.\ \cite{Chen2010}].

Further examples can be found in a variety of different areas. In the analysis of EEG data, for instance, it is of interest to locate the time point where an epileptic seizure occurs. The onset of a seizure arguably coincides with a change in the autocovariance structure of the EEG data. The aim is thus to estimate the time point where the autocovariance structure starts to vary. Another example comes from economics and concerns the Great Moderation, that is, the reduction of the volatility level of business cycle fluctuations in the middle of the 1980s. This moderation is clearly visible in the time series of U.S.\ GDP data: The volatility level of the data is roughly stable until the mid 1980s and then starts to reduce. Here, it is of interest to pin down the time point where the moderation begins. 

In most applications, there is not much known about the way in which the stochastic properties of interest evolve over time. For instance, it is not clear at all what is the functional form of the warming trend in our first example. There is no reason why it should have a particular parametric structure. Similarly, there is no economic theory suggesting that the increase of the volatility level before 2012 in our second example should have a specific parametric form. It is thus important to have flexible nonparametric methods at hand which allow to locate a smooth change point without imposing strong parametric restrictions on the time-varying behaviour of the stochastic properties under consideration.

The main goal of this paper is to develop such a method. More precisely, we tackle the following estimation problem: Suppose we observe a sample of data $\{ X_{t,T} : t=1,\ldots,T \}$ and are interested in a stochastic feature such as the mean $\ex[X_{t,T}]$ or the variance $\var(X_{t,T})$. Moreover, assume that the feature is time-invariant on the time span $\{ 1, \ldots, t_0 \}$, or equivalently, on the rescaled time interval $[0,u_0]$ with $u_0 = t_0/T$ and then starts to gradually vary over time. Our aim is to estimate the rescaled time point $u_0$. We do not impose any parametric restrictions on the time-varying behaviour of the feature of interest after $u_0$. In this sense, our model setting is completely nonparametric. Moreover, rather than restricting attention to a specific stochastic property, we set up a general procedure which allows to deal with a wide variety of features including the mean, (auto)covariances and 
higher-order  moments of the time series at hand. We tackle the problem of estimating $u_0$ within a locally stationary framework which is well suited to model gradual changes and is formally introduced in Section \ref{sec-loc-stat}.

The nonparametric nature of our estimation problem sharply distinguishes it from standard change point problems and requires new methodology. The literature commonly imposes strong parametric restrictions on the time-varying behaviour of the stochastic properties at hand. In the vast majority of papers, the changes are abrupt, that is, the properties are assumed to be constant over time apart from some occasional structural breaks. The detection of sudden structural breaks has a long history originating from statistical inference in quality control [see for example \cite{page1954,page1955} for some early references]. Since its introduction many authors have worked on this problem [see \citet{chow:1960}, \citet{brown:1975} or \citet{ploberger:1988}, among others]. Most of the literature investigates the issue of detecting breaks in the mean or the variance of a time series [see \cite{horkokste1999} or \cite{aueetal2009}], the parameters of regression models [see \cite{andrews1993} or \cite{baiper1998}] or the second order characteristics of a time series [see \cite{bergomhor2009}, \cite{wiekradeh2012}  or \cite{davis2006}]. An extensive list of references on the localization of abrupt changes can be found in \cite{jafomali2013}.

The literature on detecting gradual changes is much more scarce than that on abrupt changes. Most references consider location models of a very simple parametric form. For example, several authors investigate broken line regression models with independent normally distributed errors [see for example \cite{hinkley1970} or \cite{siezha1994}] and the performance of control charts under a gradual change in the mean [see \cite{bissel1984a,bissel1984b}  or \cite{gan1991,gan1992} among others]. Other work considers estimators and tests in models where the linear drift has been replaced by some smooth parametric  function (such as a polynomial) and the errors are assumed to be independent identically distributed but not necessarily normal [see \cite{huskova1999}, \cite{husste2002} and also \cite{aueste2002} for a generalization to the dependent case].

More recently, there has been some work on the problem of detecting smooth change points in some simple nonparametric settings. Most authors consider the location model $X_{t,T} = \mu(\frac{t}{T}) + \varepsilon_t$ with zero mean i.i.d.\ errors $\varepsilon_t$. Indeed, in many cases, the errors are even assumed to be Gaussian. Suppose that the mean function $\mu$ is constant on the interval $[0,u_0]$, i.e, $\mu(u) = \overline{\mu}$ for $u \le u_0$ and then starts to smoothly vary over time. Under appropriate smoothness conditions, $u_0$ can be regarded as a break point in the $k$-th derivative of $\mu$. It can thus be estimated by methods to detect a break point in a higher-order derivative of a nonparametric function [see \cite{Mueller1992} for an early reference and e.g.\ \cite{Raimondo1998} and \cite{Goldenshluger2006} who derive minimax rates in the model with Gaussian errors]. \cite{Mallik2011,Mallik2013} propose an alternative $p$-value based approach to estimate $u_0$ when $\mu$ is a smooth nonparametric function that is restricted to take values larger than $\overline{\mu}$ at time points $u > u_0$, that is, $\mu(u) > \overline{\mu}$ for $u > u_0$. Finally, \cite{Mercurio2004} study sequential testing procedures for change point detection in some simple nonparametric volatility models. All these methods are tailored to a very specific model setting and often rely on strong distributional assumptions. Our procedure in contrast is very general in nature and can be applied to a wide variety of settings. Moreover, it does not rely on any distributional restrictions. In the location model $X_{t,T} = \mu(\frac{t}{T}) + \varepsilon_t$, for instance, we do not even require the errors to be independent or stationary. In fact, we are able to estimate $u_0$ as long as the errors are locally stationary.

In Section \ref{sec-est}, we introduce our estimator of the time point $u_0$ which is based on a refinement of the CUSUM principle. To construct it, we proceed in two steps. In the first, we set up a function $\Dsup: [0,1] \rightarrow \reals_{\ge 0}$, where $\Dsup(u)$ measures the amount of time-variation in the stochastic feature of interest within the interval $[0,u]$. By construction, $\Dsup(u) = 0$ if there is no time-variation on the interval $[0,u]$ and $\Dsup(u) > 0$ if there is some time-variation involved. Since $\Dsup$ is not observed, we replace it by an estimator $\hat{\Dsup}_T$. Section \ref{sec-measure} gives a detailed account of how to construct the measure of time-variation $\Dsup$ and its estimator $\hat{\Dsup}_T$. The time point $u_0$ can now be characterized as the point where the measure $\Dsup$ starts to deviate from zero. This characterization is used in the second step to come up with an estimator of $u_0$. Section \ref{sec-est} describes in detail how to set up this estimator.

In Section \ref{sec-asym}, we examine the asymptotic properties of our approach. In particular, we show that the proposed estimator is consistent and derive its convergence rate. As we will see, the rate depends on the degree of smoothness of the stochastic feature of interest at $u_0$. This reflects the fact that it becomes harder to locate the time point $u_0$ when the feature varies more slowly and smoothly around this point. Our method depends on a tuning parameter with a specific statistical interpretation. In particular, it is similar in nature to a critical value in a testing procedure and can be chosen to keep a pre-specified probability of underestimating the point $u_0$. We derive a data driven choice of the tuning parameter with good theoretical and practical properties in Sections \ref{subsec-asym-threshold} and \ref{sec-impl}. The first and second part of Section \ref{sec-sim} investigate the small sample performance of our method by means of a simulation study and compare it with competing methods for the location model $X_{t,T} = \mu(\frac{t}{T}) + \varepsilon_t$. Additional simulations can be found in the Supplementary Material to the paper. Finally, in the third part of Section \ref{sec-sim}, we apply our method to the two data sets from Figure \ref{fig1}. Specifically, we use our procedure to estimate the advent of ``global warming'' and the time point prior to which the volatility level of the S\&P 500 returns strongly varies over time.

\section{Model Setting}\label{sec-loc-stat}
\def\theequation{2.\arabic{equation}}
\setcounter{equation}{0}

Throughout the paper, we assume that the sample of observations $\{ X_{t,T}: t=1,\ldots,T \}$ comes from a locally stationary process of $d$-dimensional variables $X_{t,T}$. Specifically, we work with the following concept of local stationarity, which was introduced in \cite{Vogt2012}.
\begin{definition}\label{def-loc-stat}
The array $\{ X_{t,T}: t=1,\ldots,T \}_{T=1}^{\infty}$ is called a locally stationary process if for each rescaled time point $u \in [0,1]$, there exists a strictly stationary process $\{ X_t(u): t \in \integers \}$ with the property that
\[ \bigl\| X_{t,T} - X_t(u) \bigr\| \leq \left(  \Bigl| \frac{t}{T}-u \Bigr| + \frac {1}{T} \right) U_{t,T}(u) \quad \mbox{a.s.} \]
Here, $\| \cdot \|$ denotes a norm on $\reals^d$ and $\{ U_{t,T}(u): t=1,\ldots,T \}_{T=1}^{\infty}$ is an array of positive random variables whose $\rho$-th moment is uniformly bounded for some $\rho > 0$, that is, $\ex [U^\rho_{t,T}(u)] \le C < \infty$ for some fixed constant $C$.
\end{definition}

Local stationarity was initially defined in terms of a time-varying spectral representation in  \cite{Dahlhaus1997}. Our definition of local stationarity is similar to those in \cite{Dahlhaus2006} and \cite{Koo2012} for example. The intuitive idea behind these definitions is that a process is locally stationary if it behaves approximately stationary locally in time, i.e., over short time periods. This idea is turned into a rigorous concept by requiring that locally around each rescaled time point $u$, the process $\{ X_{t,T} \}$ can be approximated by a stationary process $\{ X_t(u) \}$ in a stochastic sense.

There is a wide range of time series processes which are locally stationary in the sense of Definition \ref{def-loc-stat}. In particular, many processes with time-varying parameters can be locally approximated by a stationary process provided that the parameters are smoothly changing over time. This is fairly straightforward to show for linear models like time-varying MA or AR processes. However, it may also be verified for more complicated models like time-varying GARCH processes [see \cite{Dahlhaus2006} or \cite{SubbaRao2006}].

The definition of local stationarity relies on rescaling time to the unit interval. The main reason for doing so is to obtain a reasonable asymptotic theory. Rescaling the time argument is also common in the investigation of change points. While a completely specified parametric model as considered in \cite{hinkley1970} or \cite{siezha1994} does not need this technique, more general approaches are usually based on rescaling arguments [see \cite{huskova1999} or \cite{aueste2002} among others].

Let $\lambda_{t,T}$ be some time-varying feature of the locally stationary process $\{ X_{t,T} \}$ such as the mean $\ex[X_{t,T}]$ or the variance $\var(X_{t,T})$. Generally speaking, we allow for any feature $\lambda_{t,T}$ which fulfills the following property: 
\begin{itemize}
\item[($P_{\lambda}$)] $\lambda_{t,T}$ is uniquely determined by the set of moments $\{ \ex[f(X_{t,T})]: f \in \mathcal{F} \}$, where $\mathcal{F}$ is a family of measurable functions $f: \reals^d \rightarrow \reals$. 
\end{itemize}
Note that ($P_{\lambda}$) is a fairly weak condition which is satisfied by a wide range of stochastic features. Indeed, it essentially allows us to deal with any feature that can be expressed in terms of a set of moments. We illustrate the property ($P_{\lambda}$) by some examples:
\begin{example} \label{example1}{\rm
Let $\lambda_{t,T}$ be the mean $\mu_{t,T} = \ex[X_{t,T}]$ of a univariate locally stationary process $\{X_{t,T}\}$. Then the corresponding family of functions is simply $\mathcal{F} = \{ \text{id} \}$, since the mean $\mu_{t,T}$ can be written as $\ex[\text{id}(X_{t,T})]$.}
\end{example}
\begin{example}\label{example2}{\rm
Let $\lambda_{t,T}$ be the vector of the first $p$ autocovariances of a univariate locally stationary process $\{ Y_{t,T} \}$ whose elements $Y_{t,T}$ are centred for simplicity. Specifically, define $\gamma_{\ell,t,T} = \cov (Y_{t,T},Y_{t-\ell,T})$ to be the $\ell$-th order autocovariance and set $\lambda_{t,T} = (\gamma_{0,t,T},\ldots,\gamma_{p,t,T})^{^\intercal}$. To handle this case, we regard the data as coming from the $(p+1)$-dimensional process $\{ X_{t,T} \}$ with $X_{t,T} = (Y_{t,T},Y_{t-1,T},\ldots,Y_{t-p,T})^{^\intercal}$. We now define functions $f_{\ell}: \reals^{p+1} \rightarrow \reals$ for $0 \le \ell \le p$ by $f_{\ell}(x) = x_0 x_{\ell}$, where $x = (x_0,\ldots,x_p)^{^\intercal}$. As $\ex [f_{\ell}(X_{t,T})] = \ex [Y_{t,T} Y_{t-\ell,T}] = \gamma_{\ell,t,T}$, we obtain that $\mathcal{F} = \{ f_0,\ldots,f_p \}$ in this setting.}
\end{example}
\begin{example}\label{example3}{\rm
Consider a $d$-dimensional locally stationary process $\{X_{t,T}\}$ whose elements $X_{t,T} = (X_{t,T,1},\ldots,X_{t,T,d})^{^\intercal}$ are again centred for simplicity. Let $\lambda_{t,T}$ be the vector of covariances $\nu_{t,T}^{(i,j)} = \cov(X_{t,T,i},X_{t,T,j})$, that is, $\lambda_{t,T} = (\nu_{t,T}^{(i,j)})_{1 \le i \le j \le d}$. Analogously as in the previous example, $\mathcal{F} = \{ f_{ij} : 1 \le i \le j \le d \}$ with $f_{ij}(x) = x_i x_j$.}
\end{example}

We next define $\lambda(u)$ to be the stochastic feature of the approximating process $\{X_t(u)\}$ which corresponds to $\lambda_{t,T}$. This means that $\lambda(u)$ is fully characterized by the set of moments $\{ \ex[f(X_t(u))]: f \in \mathcal{F} \}$. Throughout the paper, we assume that 
\begin{equation}\label{gen-measure-1}
\sup_{f \in \mathcal{F}} \bigl| \ex[f(X_{t,T})] - \ex[f(X_t(u))] \bigr| \le C \Bigl( \Bigl| \frac{t}{T} - u \Bigr| + \frac{1}{T} \Bigr),
\end{equation}
which is implied by the high-order condition (C\ref{A4}) in Subsection \ref{subsec-asym-ass}. In a wide range of cases, the inequality \eqref{gen-measure-1} boils down to mild moment conditions on the random variables $X_{t,T}$, $X_t(u)$ and $U_{t,T}(u)$. This in particular holds true in Examples \ref{example1}--\ref{example3} as discussed in Subsection \ref{subsec-asym-ass}. The inequality \eqref{gen-measure-1} essentially says that $\lambda_{t,T}$ and $\lambda(u)$ are close to each other locally in time. In the time-varying mean setting from Example \ref{example1}, it can be expressed as
\[ \bigl| \mu_{t,T} - \mu(u) \bigr| \le C \Bigl( \Bigl| \frac{t}{T} - u \Bigr| + \frac{1}{T} \Bigr) \]
with $\mu(u)$ being the mean of $X_t(u)$. In Example \ref{example2}, it is equivalent to the statement
\[ \bigl\| (\gamma_{0,t,T},\ldots,\gamma_{p,t,T})^{^\intercal} - (\gamma_{0}(u),\ldots,\gamma_{p}(u))^{^\intercal} \bigr\| \le C \Bigl( \Bigl| \frac{t}{T} - u \Bigr| + \frac{1}{T} \Bigr), \]
where $\gamma_{\ell}(u) = \cov(Y_t(u),Y_{t-\ell}(u))$ and $\| \cdot \|$ is some norm on $\reals^{p+1}$. Similarly, in Example \ref{example3}, it says that
\[ \bigl\| (\nu_{t,T}^{(i.j)})_{i,j=1,\ldots,d} - (\nu^{(i,j)}(u))_{i,j=1,\ldots,d} \bigr\| \le C \Bigl( \Bigl| \frac{t}{T} - u \Bigr| + \frac{1}{T} \Bigr), \]
where $\nu^{(i,j)}(u) = \cov(X_{t,i}(u),X_{t,j}(u))$. Hence, if \eqref{gen-measure-1} holds true, the feature $\lambda_{t,T}$ converges to $\lambda(u)$ locally in time. In particular, time-variation in $\lambda_{t,T}$ is asymptotically equivalent to time-variation in $\lambda(u)$. To detect whether the stochastic feature $\lambda_{t,T}$ of interest changes over time, we may thus check for variations in the approximating quantity $\lambda(u)$.

Our estimation problem can now be formulated as follows: Assume that $\lambda(u)$ does not vary on the rescaled time interval $[0,u_0]$ but is time-varying after $u_0$. Our aim is to estimate the time point $u_0$ where $\lambda(u)$ starts to change over time.

\section{A Measure of Time-Variation}\label{sec-measure}
\def\theequation{3.\arabic{equation}}
\setcounter{equation}{0}

In this section, we set up a function $\Dsup: [0,1] \rightarrow \reals_{\ge 0}$ which captures time-variations in the stochastic feature $\lambda = \lambda(\cdot)$ of interest and explain how to estimate it. By construction, the function $\Dsup$ has the property
\begin{itemize}
\item[($P_{\Dsup}$)]{\hfil $\displaystyle{\Dsup(u)
\begin{cases}
= 0 & \text{if } \lambda \text{ does not vary on } [0,u] \\
> 0 & \text{if } \lambda \text{ varies on } [0,u]
\end{cases}}$} 
\end{itemize}
and is called a measure of time-variation. In what follows, we describe how to set up such a measure for a generic stochastic feature that satisfies ($P_{\lambda}$).

Our construction is based on the following idea: By the property ($P_{\lambda}$), the feature $\lambda(w)$ is fully characterized by the values $\ex[f(X_t(w))]$ with $f$ running over all functions in the family $\mathcal{F}$. This implies that time-variation in $\lambda(w)$ is equivalent to time-variation in the moments $\ex[f(X_t(w))]$ for some $f \in \mathcal{F}$. To detect changes in $\lambda(w)$ over time, we may thus set up a function which captures time-variations in the quantities $\ex[f(X_t(w))]$ for any $f \in \mathcal{F}$. This idea underlies the following definition:
\begin{equation}\label{Dsup-gen}
\Dsup(u) = \sup_{f \in \mathcal{F}} \sup_{v \in [0,u]} \bigl|\Dpt(u,v,f)\bigr|, 
\end{equation}
where
\begin{equation}\label{dfkt}
\Dpt(u,v,f) = \int_0^v \ex[f(X_t(w))] dw - \Bigl(\frac{v}{u}\Bigr) \int_0^u \ex[f(X_t(w))] dw. 
\end{equation}
If the moment function $\ex[f(X_t(\cdot))]$ is constant on the interval $[0,u]$, then the average $v^{-1} \int_0^v \ex[f(X_t(w))] dw$ takes the same value at all points $v \in [0,u]$. From this, it immediately follows that $D(u,v,f) = 0$ for any $v \in [0,u]$. Hence, if the function $\ex[f(X_t(\cdot))]$ is constant on $[0,u]$ for any $f \in \mathcal{F}$, then the measure of time-variation satisfies $\Dsup (u) = 0$. If $\ex[f(X_t(\cdot))]$ varies on $[0,u]$ for some $f$ in contrast, then the average $v^{-1} \int_0^v \ex[f(X_t(w))] dw$ varies on this time span as well. This is ensured by the fact that $\ex[f(X_t(\cdot))]$ is a Lipschitz continuous function of rescaled time, i.e., $|\ex[f(X_t(w))] - \ex[f(X_t(w^\prime))]| \le C|w-w^\prime|$ for any $w,w^\prime \in [0,1]$, which is a direct consequence of \eqref{gen-measure-1}. We thus obtain that $\Dpt(u,v,f) > 0$ for some $v \in [0,u]$, which in turn yields that $\Dsup(u) > 0$. As a result, $\Dsup$ satisfies ($P_{\Dsup}$).

Since the feature $\lambda$ is constant on $[0,u_0]$ but varies after $u_0$, the property ($P_{\Dsup}$) immediately implies that $\Dsup(u) = 0$ for $u \le u_0 $ and $\Dsup(u) > 0$ for $u > u_0$. The point $u_0$ is thus characterized as the time point where the measure of time-variation starts to deviate from zero. Importantly, the measure $\Dsup$ does not have a jump at $u_0$, but smoothly deviates from zero at this point. Its degree of smoothness depends on how smoothly the moments $\ex[f(X_t(w))]$ vary over time, or put differently, on how smoothly the feature $\lambda(w)$ varies over time. In particular, the smoother the time-variation in $\lambda$, the smoother the function $\Dsup$.

In order to estimate the measure of time-variation, we proceed as follows: The integral $\int_0^v \ex[f(X_t(w))] dw$ can be regarded as an average of the moments $\ex[f(X_t(w))]$, where all time points from $0$ to $v$ are taken into account. This suggests to estimate it by a sample average of the form $T^{-1} \sum\nolimits_{t=1}^{\lfloor vT \rfloor} f(X_{t,T})$. Following this idea, an estimator of $\Dsup(u)$ is given by 
\[ \hat{\Dsup}_T(u) = \sup_{f \in \mathcal{F}} \sup_{v \in [0,u]} \bigl|\hat{\Dpt}_T(u,v,f)\bigr|, \]
where we set
\[ \hat{\Dpt}_T(u,v,f) = \frac{1}{T} \sum\limits_{t = 1}^{\lfloor vT \rfloor} f(X_{t,T}) - \Bigl(\frac{v}{u}\Bigr) \frac{1}{T} \sum\limits_{t = 1}^{\lfloor uT \rfloor} f(X_{t,T}). \]
The statistic $\hat{\Dsup}_T(u)$ is constructed by the CUSUM principle for the interval $[0,u]$ and can be regarded as a generalization of classical CUSUM statistics to be found for example in \cite{page1954,page1955}. The quantity $\hat{\Dpt}_T$ compares cumulative sums of the variables $f(X_{t,T})$ over different time spans $[0,v]$ and $[0,u]$. By taking the supremum with respect to $v \in [0,u]$, we are able to detect gradual changes in the signal $\ex[f(X_t(\cdot)]$ on the interval $[0,u]$. The additional supremum over $f$ makes sure that the signals corresponding to all functions $f \in \mathcal{F}$ are taken into account.

\section{Estimating the Gradual Change Point $\boldsymbol{u_0}$}\label{sec-est}
\def\theequation{4.\arabic{equation}}
\setcounter{equation}{0}

We now describe how to use our measure of time-variation to estimate the point $u_0$. Our estimation method is based on the observation that $\sqrt{T} \Dsup(u) = 0$ for $u \le u_0 $ and $\sqrt{T} \Dsup(u) \rightarrow \infty$ for $u > u_0$ as $T \rightarrow \infty$. The scaled estimator $\sqrt{T} \hat{\Dsup}_T(u)$ behaves in a similar way: As we will see later on, 
\begin{equation}\label{Dsup-conv}
\sqrt{T} \hat{\Dsup}_T(u)
\begin{cases}
\convd \Hsup(u) & \text{for } u \le u_0 \\
\convp \infty   & \text{for } u > u_0,
\end{cases}
\end{equation}
where $\Hsup(u)$ is a real-valued random variable. By \eqref{Dsup-conv}, $\sqrt{T} \hat{\Dsup}_T(u)$ can be regarded as a statistic to test the hypothesis that the feature of interest $\lambda$ is time-invariant on the interval $[0,u]$. Under the null of time-invariance, that is, as long as $u \le u_0$, the statistic weakly converges to some limit distribution. Under the alternative, that is, at time points $u > u_0$, it diverges in probability to infinity. The main idea of the new  estimation method is to exploit this dichotomous behaviour.

To construct our estimator of $u_0$, we proceed as follows: First of all, we define the quantity
\[ \hat{r}_T(u) = 1(\sqrt{T} \hat{\Dsup}_T(u) \le \tau_T), \]
where $\tau_T$ is a threshold level that slowly diverges to infinity. A data driven choice of $\tau_T$ with good theoretical and practical properties is discussed in detail in Section \ref{subsec-asym-threshold}. The random variable $\hat{r}_T(u)$ specifies the outcome of our test on time-invariance for the interval $[0,u]$ given the critical value $\tau_T$: if the test accepts the null of time-invariance, then $\hat{r}_T(u) = 1$; if it rejects the null, then $\hat{r}_T(u) = 0$. Under the null, the test statistic tends to take moderate values, suggesting that $\hat{r}_T(u)$ should eventually become zero. Under the alternative, the statistic explodes, implying that $\hat{r}_T(u)$ should finally take the value one. Formally speaking, one can show that  
\[ \hat{r}_T(u) \convp
\begin{cases}
1  & \text{for } u \le u_0 \\
0  & \text{for } u > u_0,
\end{cases}
\]
if $\tau_T$ converges (slowly)  to infinity. This suggests that $\int_0^1 \hat{r}_T(u) du \approx u_0$ for large sample sizes. Hence, we may simply estimate $u_0$ by aggregating the test outcomes $\hat{r}_T(u)$, that is, 
\[ \hat{u}_0(\tau_T) = \int_0^1 \hat{r}_T(u) du. \]
This estimator exploits the fact that the test outcome should be equal to one at time points $u \le u_0$ but equal to zero at $u > u_0$.

\section{Asymptotic Properties}\label{sec-asym}
\def\theequation{5.\arabic{equation}}
\setcounter{equation}{0}

We now examine the asymptotic properties of the proposed  estimation method. We first investigate the weak convergence behaviour of the statistic $\hat{\Dpt}_T$ and then derive the convergence rate of the estimator $\hat{u}_0(\tau_T)$. Since the proofs are very technical and involved, they are deferred to the Appendix. To state the results, we let the symbol $\ell_{\infty}(S)$ denote the space of bounded functions $f: S \rightarrow \reals$ endowed with the supremum norm and let $\convw$ denote weak convergence. Moreover, to capture the amount of smoothness of the measure $\Dsup$ at the point $u_0$, we suppose that
\begin{equation}\label{smoothness-D}
\frac{\Dsup(u)}{(u_0 - u)^{\kappa}} \rightarrow c_{\kappa} > 0 \qquad \text{as } u \searrow u_0
\end{equation}
for some number $\kappa > 0$ and a constant $c_{\kappa} > 0$. The larger $\kappa$, the more smoothly the measure $\Dsup$ deviates from zero at the point $u_0$.

\subsection{Assumptions}\label{subsec-asym-ass}

Throughout the paper, we make the following assumptions:
\begin{listing}{C}{}
\item \label{A1} The process $\{X_{t,T}\}$ is locally stationary in the sense of Definition \ref{def-loc-stat}.
\item \label{A2} The process $\{X_{t,T}\}$ is strongly mixing with mixing coefficients $\alpha(k)$ satisfying $\alpha(k) \le C a^k$ for some positive constants $C$ and $a < 1$.
\item \label{A3} Let $p \ge 4$ be an even natural number and endow the set $\mathcal{F}$ with some semimetric $\dist$. $(\mathcal{F},\dist)$ is separable, compact and not too complex in the sense that its covering number $\mathcal{N}(w,\mathcal{F},\dist)$ satisfies the condition
$\int_0^1 \mathcal{N}(w,\mathcal{F},\dist)^{1/p} dw  < \infty$.
Moreover, the set $\mathcal{F}$ has an envelope $F$ (i.e.\ $|f| \le F$ for all $f \in \mathcal{F}$) which satisfies $\ex [F(X_{t,T})^{(1+\delta)p}] \le C < \infty$ for some small $\delta > 0$ and a fixed constant $C$. Finally, for any pair of functions $f,f^{\prime} \in \mathcal{F}$,
\[ \ex \Bigl[ \Bigl| \frac{f(X_{t,T}) - f^{\prime}(X_{t,T})}{\dist(f,f^{\prime})} \Bigr|^{(1+\delta)p} \Bigr] \le C < \infty. \]
\item \label{A4} For $k=1,2$ and all $f \in \mathcal{F}$, it holds that $\ex [|f(X_{t,T}) - f(X_t(u))|^k] \le C (|\frac{t}{T}-u| + \frac{1}{T})$ for some fixed constant $C$.
\end{listing}
Condition (C\ref{A2}) stipulates that the array $\{X_{t,T}\}$ is strongly mixing. A wide variety of locally stationary processes can be shown to be mixing under appropriate conditions; see for example \cite{SubbaRao2011} and  \cite{Vogt2012}. To keep the structure of the proofs as clear as possible, we have assumed the mixing rates to decay exponentially fast. Alternatively, we could work with slower polynomial rates at the cost of a more involved notation in the proofs. Conditions (C\ref{A3}) and (C\ref{A4}) allow for a wide range of function families $\mathcal{F}$ and are formulated in a very general way. For many choices of $\mathcal{F}$, they boil down to simple moment conditions on the variables $X_{t,T}$, $X_t(u)$ and $U_{t,T}(u)$. We illustrate this by means of Examples \ref{example1}--\ref{example3}. It is straightforward to show that in Example \ref{example1}, (C\ref{A3}) and (C\ref{A4}) are satisfied under the following set of moment conditions:
\begin{itemize}
\item[$(A_{\mu})$] Either (a) $\ex |X_{t,T}|^r \le C$ for some $r > 4$ and $\ex U_{t,T}^2(u) \le C$ or (b) $\ex |X_{t,T}|^r \le C$, $\ex |X_t(u)|^r \le C$ and $\ex U_{t,T}^{r/(r-1)}(u) \le C$ for some $r > 4$ and a sufficiently large constant $C$ that is independent of $u$, $t$ and $T$.
\end{itemize}
Similarly, in Example \ref{example2}, they are implied by 
\begin{itemize}
\item[$(A_{\gamma})$] $\ex \|X_{t,T}\|^r \le C$, $\ex \|X_t(u)\|^r \le C$ and $\ex U_{t,T}^q(u) \le C$ for some $r > 8$ and $q = \frac{r}{3} / (\frac{r}{3}-1)$, where $C$ is a sufficiently large constant that is independent of $u$, $t$ and $T$.
\end{itemize}
The moment conditions in Example \ref{example3} are fully analogous to those in Example \ref{example2} and thus not stated explicitly. 

\subsection{Weak convergence of the measure of time-variation}\label{subsec-asym-measure}

To start with, we investigate the asymptotic properties of the expression
\begin{equation}\label{Hpt}
\hat{\Hpt}_T(u,v,f) = \sqrt{T} \bigl( \hat{\Dpt}_T(u,v,f) - \Dpt(u,v,f) \bigr). 
\end{equation}
To do so, let $\Delta = \{(u,v) \in [0,1]^2: v \le u \}$ and equip the space $\Delta \times \mathcal{F}$ with the natural semimetric $|u - u^\prime| + |v - v^\prime| + \dist(f,f^\prime)$. In what follows, we regard $\hat{\Hpt}_T$ as a process that takes values in $\ell_{\infty}(\Delta \times \mathcal{F})$ and show that it weakly converges to a Gaussian process $\Hpt$ with the covariance structure
\begin{align}
\cov(\Hpt(u,v,f),\Hpt(u^\prime,v^\prime,f^\prime)) 
 & = \sum\limits_{l=-\infty}^{\infty} \Bigl\{ \frac{v v^\prime}{u u^\prime} \int_0^{\min \{u,u^\prime\}} c_l(w) dw - \frac{v^\prime}{u^\prime} \int_0^{\min \{v,u^\prime\}} c_l(w) dw \nonumber \\
 & \phantom{= \sum\limits_{l=-\infty}^{\infty} \Bigl\{ } - \frac{v}{u} \int_0^{\min \{u,v^\prime\}} c_l(w) dw + \int_0^{\min \{v,v^\prime\}} c_l(w) dw \Bigr\}, \label{cov-H}
\end{align}
where $c_l(w) = c_l(w,f,f^{\prime}) = \cov(f(X_0(w)),f^\prime(X_l(w)))$. The following theorem gives a precise description of the weak convergence of $\hat{\Hpt}_T$.
\begin{theorem}\label{theo-measure}
Let assumptions (C\ref{A1})--(C\ref{A4}) be satisfied. Then
\[ \hat{\Hpt}_T = \sqrt{T} \bigl[ \hat{\Dpt}_T - \Dpt \bigr] \convw \Hpt \]
as a process in $\ell_{\infty}(\Delta \times \mathcal{F})$, where $\hat{\Dpt}_T$ and $\Dpt$ are defined in Section \ref{sec-measure} and $H$ is a Gaussian process on $\Delta \times \mathcal{F}$  with covariance kernel \eqref{cov-H}.
\end{theorem}
This theorem is the main stepping stone to derive the asymptotic properties of our estimator $\hat{u}_0(\tau_T)$. In addition, it is useful to examine the asymptotic behaviour of some processes related to $\hat{\Hpt}_T$: Analogously to $\hat{\Dsup}_T(u)$, we introduce the expression
\begin{equation}\label{Hsup}
\hat{\Hsup}_T(u) = \sup_{f \in \mathcal{F}} \sup_{v \in [0,u]} \bigl| \hat{\Hpt}_T(u,v,f) \bigr|.  
\end{equation}
Moreover, we let
\begin{equation}\label{Dmax}
\hat{\Dmax}_T(u) = \sup_{v \in [0,u]} \hat{\Dsup}_T(v) = \sup_{f \in \mathcal{F}} \sup_{0 \le w \le v \le u} \bigl| \hat{\Dpt}_T(v,w,f) \bigr| 
\end{equation}
together with
\begin{equation}\label{Hmax}
\hat{\Hmax}_T(u) = \sup_{v \in [0,u]} \hat{\Hsup}_T(v) = \sup_{f \in \mathcal{F}} \sup_{0 \le w \le v \le u} \bigl| \hat{\Hpt}_T(v,w,f) \bigr|. 
\end{equation}
The next result directly follows from Theorem \ref{theo-measure} together with the continuous mapping theorem.
\begin{corollary}\label{corollary-measure}
Let assumptions (C\ref{A1})--(C\ref{A4}) be satisfied. Then
\[ \hat{\Hsup}_T \convw \Hsup \quad \text{and} \quad \hat{\Hmax}_T \convw \Hmax \] 
as processes in $\ell_{\infty}([0,1])$, where  $\Hsup$ and $\Hmax$ are defined by $\Hsup(u) = \sup_{f \in \mathcal{F}, v \in [0,u]} |\Hpt(u,v,f)|$ and $\Hmax(u) = \sup_{f \in \mathcal{F}, 0 \le w \le v \le u} |\Hpt(v,w,f)|$, respectively.
\end{corollary}

\subsection{Convergence of the estimator $\boldsymbol{\hat{u}_0(\tau_T)}$}\label{subsec-asym-convergence}

We now turn to the asymptotic behaviour of the estimator $\hat{u}_0(\tau_T)$. The next theorem shows that $\hat{u}_0(\tau_T)$ consistently estimates $u_0$ provided that the threshold level $\tau_T$ diverges to infinity. Moreover, it specifies the convergence rate at which $\hat{u}_0(\tau_T)$ approaches $u_0$.
\begin{theorem}\label{theo-convergence}
Let assumptions (C\ref{A1})--(C\ref{A4}) be satisfied and assume that $\tau_T \rightarrow \infty$ with $\tau_T / \sqrt{T} \rightarrow 0$. Then
\[ \hat{u}_0(\tau_T) - u_0 = O_p(\gamma_T), \]
where $\gamma_T = (\tau_T / \sqrt{T})^{1/\kappa}$ and $\kappa$ is defined in \eqref{smoothness-D}.
\end{theorem}
The convergence rate of $\hat{u}_0(\tau_T)$ can be seen to depend on the degree of smoothness $\kappa$ of the measure $\Dsup$ at the point $u_0$. In particular, the smoother $\Dsup$, the slower the convergence rate. Since the smoothness of $\Dsup$ mirrors that of the stochastic feature $\lambda$, we can equivalently say: the smoother the feature $\lambda$ varies around $u_0$, the slower the rate of our estimator gets. This reflects the intuition that it becomes harder to precisely localize the point $u_0$ when $\lambda$ varies more smoothly and gradually around this point. The rate $\gamma_T$ also depends on the threshold parameter $\tau_T$. Specifically, the slower $\tau_T$ diverges to infinity, the faster the rate $\gamma_T$ goes to zero. Hence, from a theoretical point of view, $\tau_T$ should be chosen to diverge as slowly as possible to speed up the convergence rate of the estimator.

\subsection{Choice of the threshold level $\boldsymbol{\tau_T}$}\label{subsec-asym-threshold}

We next discuss how to choose the threshold $\tau_T$ to obtain an estimator of $u_0$ with good theoretical properties. To state the results, we let $q_{\alpha}(u)$ be the $(1-\alpha)$-quantile of the limiting variable $\Hmax(u)$ and assume throughout that this quantile is known for any time point $u$. In practice, it is indeed unknown and has to be approximated. We show how to achieve this in Section \ref{sec-impl} where we discuss the implementation of our method. Our choice of the threshold $\tau_T$ proceeds in two steps. In the first, we describe a rough choice of $\tau_T$ which leads to a preliminary estimator of $u_0$. In the second, we use this preliminary estimator to come up with a refined choice of $\tau_T$ which in turn yields a better estimator of $u_0$. 

\vspace{10pt}

\textbf{Preliminary choice of $\boldsymbol{\tau_T}$.} To convey the idea behind the choice of $\tau_T$, let us first assume that $\tau_T$ does not depend on the sample size, i.e., $\tau_T = \tau$ for all $T$. A first crude choice of $\tau$ can be obtained by arguing in a similar way as in classical change point detection problems: Consider the situation that the stochastic feature of interest is time-invariant on $[0,1]$, i.e., there is no change point $u_0 < 1$. In this situation, we would like to control the probability of false detection of a change point. Specifically, we aim to choose $\tau$ such that this probability is smaller than some pre-specified level $\alpha$, that is,
\[ \pr( \hat{u}_0(\tau) < 1) \le \alpha, \]
when there is no change point $u_0 < 1$. To achieve this, we write
\begin{align*}
\pr (\hat{u}_0(\tau) < 1) 
 & \le \pr \bigl( \sqrt{T} \hat{\Dsup}_T(u) > \tau \text{ for some } u \in [0,1] \bigr) 
   = \pr \bigl( \sqrt{T} \hat{\Dmax}_T(1) > \tau \bigr).
\end{align*}
Corollary \ref{corollary-measure} shows that $\sqrt{T} \hat{\Dmax}_T(u)$ weakly converges to the limiting variable $\Hmax(u)$ at each point $u \le u_0$. In particular, when there is no change point $u_0 < 1$, it holds that $\sqrt{T} \hat{\Dmax}_T(1) \convd \Hmax(1)$. We now set $\tau$ to be the $(1-\alpha)$-quantile $q_{\alpha}(1)$ of $\Hmax(1)$. Writing $\tau_{\alpha}^{\circ} = q_{\alpha}(1)$, we obtain that
\[ \pr (\hat{u}_0(\tau_{\alpha}^{\circ}) < 1) \le \alpha + o(1), \]
when there is no change point $u_0 < 1$. We are thus able to asymptotically control the probability of false detection by choosing $\tau = \tau_{\alpha}^{\circ}$. However, this choice does not yield a consistent estimator of $u_0$. To ensure consistency, we have to make sure that the threshold $\tau_T$ diverges to infinity. To achieve this, we let the level $\alpha_T$ depend on the sample size $T$. In particular, we let it slowly converge to zero and set $\tau_T = \tau_{\alpha_T}^{\circ}$. 

\vspace{10pt}

\textbf{Refined choice of $\boldsymbol{\tau_T}$.} As in classical change point problems, the choice $\tau = \tau_{\alpha}^{\circ}$ is fairly conservative. In particular, the resulting estimator tends to strongly overestimate the time point $u_0$. In what follows, we refine the choice of $\tau$ to get a more precise estimator of $u_0$. Rather than controlling the false detection rate, we would like to control the probability of underestimating $u_0$, i.e., of falsely detecting a change point before it occurs. Technically speaking, we aim to choose $\tau$ such that
\[ \pr(\hat{u}_0(\tau) < u_0) \le \alpha \]
for some given level $\alpha$. Similarly as above, it holds that
\begin{align*}
\pr (\hat{u}_0(\tau) < u_0) 
 & \le \pr \bigl( \sqrt{T} \hat{\Dsup}_T(u) > \tau \text{ for some } u \in [0,u_0] \bigr)
   = \pr \bigl( \sqrt{T} \hat{\Dmax}_T(u_0) > \tau).
\end{align*}
By Corollary \ref{corollary-measure}, we know that $\sqrt{T} \hat{\Dmax}_T(u_0) \convd \Hmax(u_0)$. Setting $\tau$ to equal the $(1-\alpha)$-quantile $q_{\alpha}(u_0)$ of the limiting variable $\Hmax(u_0)$ and using the notation $\tau_{\alpha} = q_{\alpha}(u_0)$, we are able to derive the following result.
\begin{theorem}\label{theo-threshold}
Let assumptions (C\ref{A1})--(C\ref{A4}) be satisfied. Then
\begin{equation}\label{error-underest}
\pr(\hat{u}_0(\tau_{\alpha}) < u_0) \le \alpha + o(1) 
\end{equation}
and for any constant $C > 0$,
\begin{equation}\label{error-overest}
\pr(\hat{u}_0(\tau_{\alpha}) > u_0 + C \gamma_T) = o(1),
\end{equation}
where $\gamma_T$ is defined in Theorem \ref{theo-convergence}.
\end{theorem}
Hence, the estimator $\hat{u}_0(\tau_{\alpha})$ has the following properties: According to \eqref{error-underest}, the probability of underestimating $u_0$ is asymptotically bounded by $\alpha$. Moreover, the probability of overestimating $u_0$ by more than $C \gamma_T$ is asymptotically negligible by \eqref{error-overest}. Thus, $\hat{u}_0(\tau_{\alpha})$ controls the error of underestimating $u_0$ while being consistent when it comes to overestimation.

Of course, we cannot take the choice $\tau = \tau_{\alpha}$ at face value since the quantile $\tau_{\alpha} = q_{\alpha}(u_0)$ depends on the unknown location $u_0$. Nevertheless, we can estimate this quantile by $\hat{\tau}_{\alpha} = q_{\alpha}(\hat{u}_0(\tau_T^{\circ}))$, where $\hat{u}_0(\tau_T^{\circ})$ is a consistent pilot estimator of $u_0$. In particular, we may set $\tau_T^{\circ} = \tau_{\alpha_T}^{\circ}$ and use $\hat{u}_0(\tau_{\alpha_T}^{\circ})$ as a pilot estimate. It is fairly straightforward to see that the statements of Theorem \ref{theo-threshold} still hold true when $\tau_{\alpha}$ is replaced by $\hat{\tau}_{\alpha}$:
\begin{corollary}\label{corollary-threshold}
Let assumptions (C\ref{A1})--(C\ref{A4}) be satisfied. Then
\begin{equation}\label{error-underest-cor}
\pr(\hat{u}_0(\hat{\tau}_{\alpha}) < u_0) \le \alpha + o(1) 
\end{equation}
and for any $C > 0$,
\begin{equation}\label{error-overest-cor}
\pr(\hat{u}_0(\hat{\tau}_{\alpha}) > u_0 + C \gamma_T) = o(1). 
\end{equation}
\end{corollary}
As in the previous subsection, we suggest to set $\tau_T = \hat{\tau}_{\alpha_T}$ with $\alpha_T$ gradually converging to zero  to obtain a consistent estimator of $u_0$.

\section{Implementation}\label{sec-impl}
\def\theequation{6.\arabic{equation}}
\setcounter{equation}{0}

To implement our estimation method in practice, we proceed as follows:
\begin{enumerate}[leftmargin=2.25cm,label=\textit{Step \arabic*.},itemsep=0cm]
\item Fix a probability level $\alpha$ and estimate the threshold parameter $\tau_\alpha$. \\[-0.85cm]
\begin{enumerate}[leftmargin=0.85cm,label=(\roman*),itemsep=0cm]
\item Approximate the quantiles $q_{\alpha}(u)$ by $\hat{q}_{\alpha}(u)$ as described below.   
\item Compute the preliminary estimator $\hat{u}_0^{\circ} = \hat{u}_0(\hat{\tau}_{\alpha}^{\circ})$, where $\hat{\tau}_{\alpha}^{\circ} = \hat{q}_{\alpha}(1)$.
\item Estimate $\tau_{\alpha}$ by $\hat{\tau}_{\alpha} = \hat{q}_{\alpha}(\hat{u}_0^{\circ})$. \\[-0.85cm]
\end{enumerate}
\item Estimate $u_0$ by $\hat{u}_0(\hat{\tau}_{\alpha})$.
\end{enumerate}

Generally speaking, the quantiles $q_{\alpha}(u)$ can be approximated as follows: By definition,
\[ \Hmax(u) = \sup_{f \in \mathcal{F}} \sup_{0 \le w \le v \le u} \bigl| \Hpt(v,w,f) \bigr| \]
is the supremum of the Gaussian process $\Hpt$ whose covariance structure is given in \eqref{cov-H}. Inspecting the formula \eqref{cov-H}, the only unknown expressions occurring in it are of the form 
\[ \sigma^2(u,f,f^{\prime}) = \sum\limits_{l=-\infty}^{\infty} \Gamma_l(u,f,f^{\prime}), \]
where $\Gamma_l(u,f,f^{\prime}) = \int_0^u c_l(w,f,f^{\prime}) dw$ and $c_l(w,f,f^{\prime}) = \cov(f(X_0(w)),f^{\prime}(X_l(w)))$. These quantities are essentially average long-term covariances of the processes $\{f(X_t(w))\}$ and $\{f^{\prime}(X_t(w))\}$ on the interval $[0,u]$, which can be estimated by methods for long-run covariance estimation. Specifically, we can employ HAC type estimation procedures as discussed in \cite{Andrews1991} or \cite{deJong2000} among others and work with an estimator of the form 
\begin{equation}\label{est-lrv}
\hat{\sigma}^2(u,f,f^{\prime}) = \sum\limits_{l=-\infty}^{\infty} K\Bigl(\frac{l}{b(T)}\Bigr) \hat{\Gamma}_l(u,f,f^{\prime}). 
\end{equation}
where $K$ is a kernel of Bartlett or flat-top type and $b=b(T)$ is the bandwidth. Moreover, 
\[ \hat{\Gamma}_l(u,f,f^{\prime}) = \frac{1}{T} \sum\limits_{t=1}^{\lfloor uT \rfloor} \hat{Z}_{t,T}(f) \hat{Z}_{t-l,T}(f^{\prime}), \]
where $\hat{Z}_{t,T}(f) = f(X_{t,T}) - \hat{\ex}[f(X_{t,T})]$ and $\hat{\ex}[f(X_{t,T})]$ is an estimator of $\ex[f(X_{t,T})]$. We may for example use a Nadaraya-Watson estimate
\[ \hat{\ex}[f(X_{t,T})] = \frac{1}{T} \sum_{s=1}^T K_h \Bigl( \frac{t}{T} - \frac{s}{T} \Bigr) f(X_{s,T}) \]
with $K$ being a kernel function and $K_h(x) = h^{-1} K(x/h)$. Alternatively, a local linear or more generally a local polynomial smoother may be employed. Once we have calculated the estimator $\hat{\sigma}^2(u,f,f^{\prime})$, we can compute the covariance function \eqref{cov-H} and simulate observations from the Gaussian process with the estimated covariance structure. This in turn allows us to simulate the quantiles $q_{\alpha}(u)$.

Our implementation strategy works well in practice as we will demonstrate in the empirical part of the paper. When the class of functions $\mathcal{F}$ is large, it becomes computationally more burdensome to simulate the quantiles $q_{\alpha}(u)$. In most applications, however, the class of functions is fairly small. Moreover, in a number of cases, it is possible to simplify the implementation by exploiting the special structure of the model at hand. To illustrate this, we revisit the simple time-varying mean setting from Example \ref{example1}. In particular, we consider the model
\begin{equation}\label{mod0}
X_{t,T} = \mu \Bigl( \frac{t}{T} \Bigr) + \varepsilon_{t,T}, 
\end{equation}
where the feature $\lambda_{t,T}$ of interest is given by the mean function $\mu(\frac{t}{T})$. Recalling that $\mathcal{F} = \{ \text{id} \}$ in this case, the covariance structure \eqref{cov-H} depends on the expressions $\sigma^2(u) = \sum\nolimits_{l=-\infty}^{\infty} \int_0^u c_l(w) dw$, where $c_l(w) = \ex[\varepsilon_0(w) \varepsilon_l(w)]$ and $\{\varepsilon_t(w)\}$ is the stationary approximating process of $\{\varepsilon_{t,T}\}$ at the time point $w$. If the error process is stationary, we even obtain that $\sigma^2(u) = u \sigma^2$ for all $u$, where $\sigma^2 = \sum\nolimits_{l=-\infty}^{\infty} \ex[ \varepsilon_0 \varepsilon_l]$ is the long-run variance of the error terms. The latter can be estimated by standard methods. Denoting its estimator by $\hat{\sigma}^2$, we can set up our method in terms of the scaled statistic $\hat{\Dsup}_T^{\text{sc}}(u) = \hat{\Dsup}_T(u)/ \hat{\sigma}$. Defining the expressions $\hat{\Dpt}_T^{\text{sc}}(u)$, $\hat{\Hpt}_T^{\text{sc}}(u)$ etc.\ in an analogous way, we obtain that $\hat{\Hpt}_T^{\text{sc}} \convw \Hpt^{\text{sc}}$, where the Gaussian process $\Hpt^{\text{sc}}$ has the covariance structure
\[ \cov(\Hpt^{\text{sc}}(u,v),\Hpt^{\text{sc}}(u^\prime,v^\prime)) = \frac{v v^\prime}{u u^\prime} \min \{u,u^\prime\} - \frac{v^\prime}{u^\prime} \min \{v,u^\prime\} - \frac{v}{u} \min \{u,v^\prime\} + \min \{v,v^\prime\}. \]
Importantly, this formula does not involve any unknown quantities, which in turn means that the quantiles $q^{\text{sc}}_{\alpha}(u)$ of $\Hmax^{\text{sc}}(u)$ are completely known (neglecting the simulation error). Consequently, in this setting, which is often of interest in statistical practice, the method is particularly easy to implement.

\section{Finite Sample Properties}\label{sec-sim}

\subsection{Simulations}\label{subsec-sim}
\def\theequation{7.\arabic{equation}}
\setcounter{equation}{0}

In this and the following subsection, we examine the small sample performance of our estimation method in a Monte-Carlo experiment. We first investigate two simulation setups which are motivated by the applications in the introduction: a time-varying mean model and a volatility model together with a multivariate extension of it. Due to space constraints, the results on the volatility models are presented in the Supplementary Material. Here, we examine the setting
\begin{equation} \label{mod1} 
X_{t,T} = \mu \Bigl( \frac{t}{T} \Bigr) + \varepsilon_t 
\end{equation}
with two different mean functions  $\mu_1$ and $\mu_2$. The residuals $\varepsilon_t$ are assumed to follow the AR(1) process $\varepsilon_t = 0.25 \varepsilon_{t-1} + \eta_t$, where the innovations $\eta_t$ are i.i.d.\ normal with zero mean and standard deviation $0.5$. The mean functions are given by 
\begin{align}
\mu_1(u) & = 1(u > 0.5) \label{mu1a} \\
\mu_2(u) & = \{10(u-0.5)\} \cdot 1(0.5 < u < 0.6) + 1(u > 0.6). \label{mu2a} 
\end{align}
Both functions are equal to zero on the interval $[0,0.5]$ and then start to vary over time. Hence, $u_0 = 0.5$ in both cases. The function $\mu_1$ is a step function which allows to investigate how our method works in the presence of abrupt changes. The function $\mu_2$ in contrast varies smoothly over time. In particular, it starts to linearly deviate from zero at the point $u_0 = 0.5$ until it reaches a value of one and then remains constant.

To estimate the point $u_0$, we use the implementation strategy from Section \ref{sec-impl} and denote the resulting estimator by $\hat{u}_0$. We set the parameter $\alpha$ to equal $0.1$ in all our simulations, meaning that the probability of underestimating $u_0$ is approximately $10\%$. Moreover, as described at the end of Section \ref{sec-impl}, we normalize the statistic $\hat{\Dsup}_T(u)$ by an estimate of the long-run error variance $\sigma^2 = \sum\nolimits_{l=-\infty}^{\infty} \ex[ \varepsilon_0 \varepsilon_l]$. To do so, we first approximate the residuals $\varepsilon_t$ by $\hat{\varepsilon}_t = X_{t,T} - \hat{\mu}_h(\frac{t}{T})$, where $\hat{\mu}_h$ is a Nadaraya-Watson estimator of $\mu$, and then apply a HAC estimator with a Bartlett kernel to the estimated residuals. Here, we set $h = 0.2$ and choose the bandwidth of the HAC estimator to equal $10$, i.e., we take into account the first ten autocovariances. As a robustness check, we have repeated the simulations for different bandwidth parameters. As this yields very similar results, we have not reported them here.

\begin{figure}[H]
\centering
\includegraphics[width=0.32\textwidth]{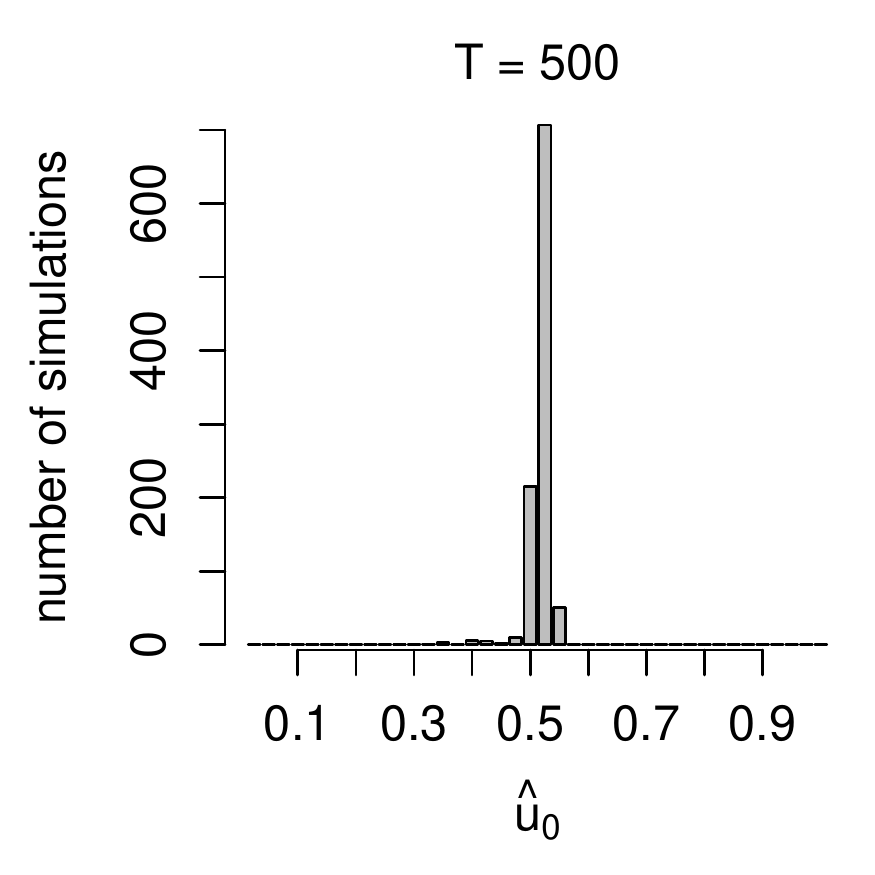}
\includegraphics[width=0.32\textwidth]{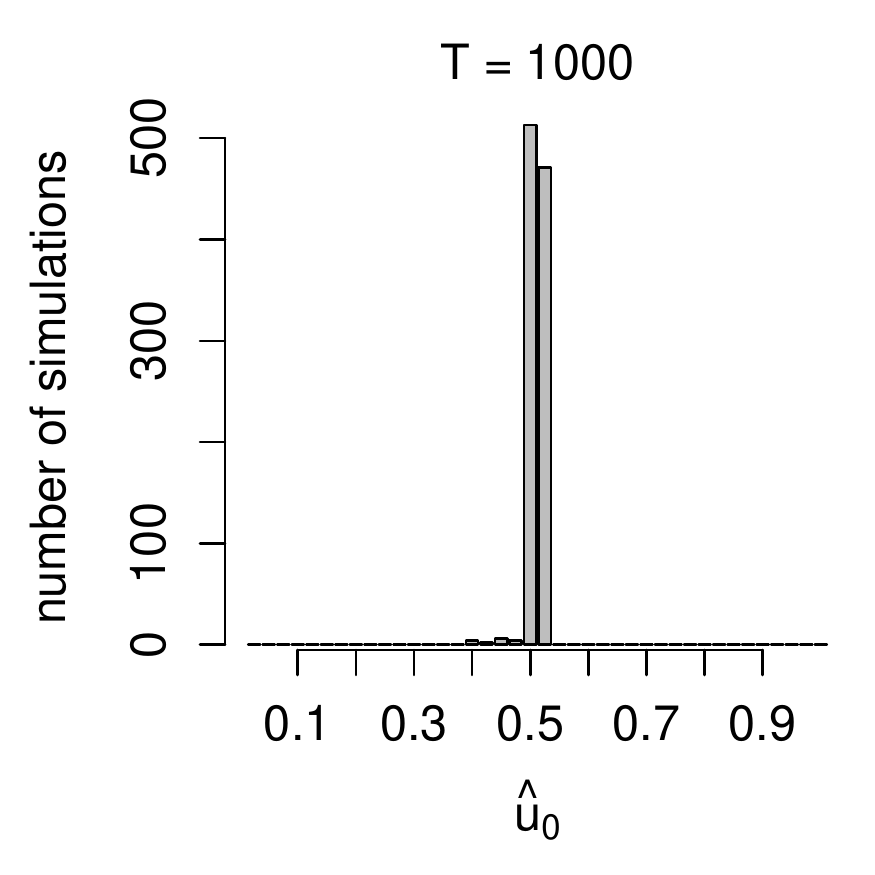} \\
\smallskip
\includegraphics[width=0.32\textwidth]{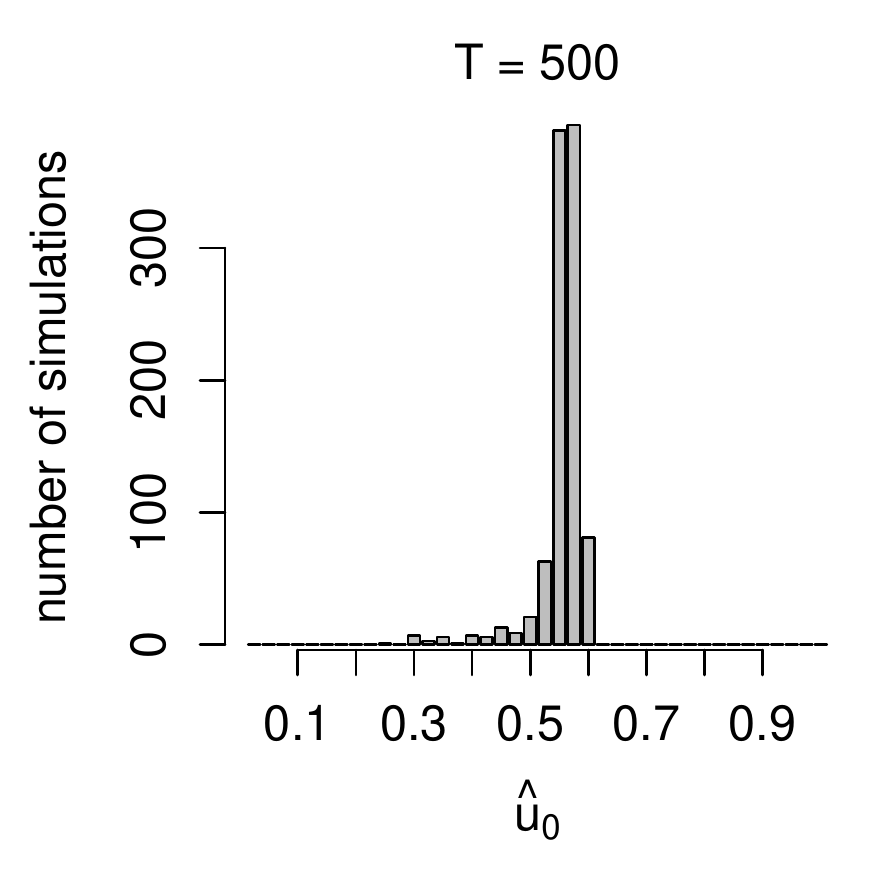}
\includegraphics[width=0.32\textwidth]{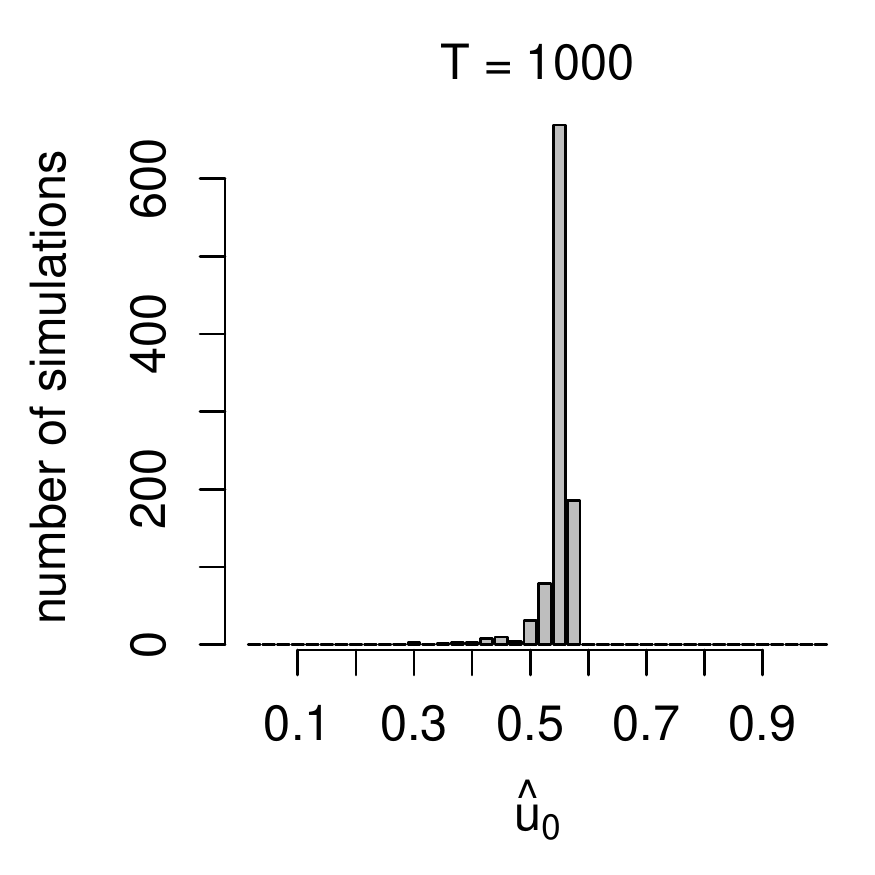}
\caption{Simulation results produced by our method in model \eqref{mod1}. Upper panel: results for the mean function $\mu_1$ defined in \eqref{mu1a}. Lower panel: results for the mean function $\mu_2$ defined in \eqref{mu2a}.}\label{fig-mu}
\end{figure}

For each model setting, we produce $N = 1000$ samples of length $T \in \{ 500, 1000 \}$ and apply our procedure to estimate $u_0$. We thus obtain $N=1000$ estimates of $u_0$ for each model specification. The  results are presented by histograms that show the empirical distribution of the estimates for each specification. In particular, the bars in the plots give the number of simulations (out of a total of $1000$) in which a certain value $\hat{u}_0$ is obtained.

The simulation results for the design with $\mu_1$ are presented in the upper part of Figure \ref{fig-mu}, the left-hand panel corresponding to a sample size of $T=500$ and the right-hand one to $T=1000$. Since $\mu_1$ has a jump at $u_0 = 0.5$, it deviates from zero very quickly. Our procedure is thus able to localize the point $u_0$ quite precisely. This becomes visible in the histograms which show that the estimates are not very dispersed but cluster tightly around $u_0=0.5$. The plots also make visible a slight upward bias which becomes less pronounced when moving to the larger sample size $T=1000$.

\begin{figure}[H]
\centering
\includegraphics[width=0.6\textwidth]{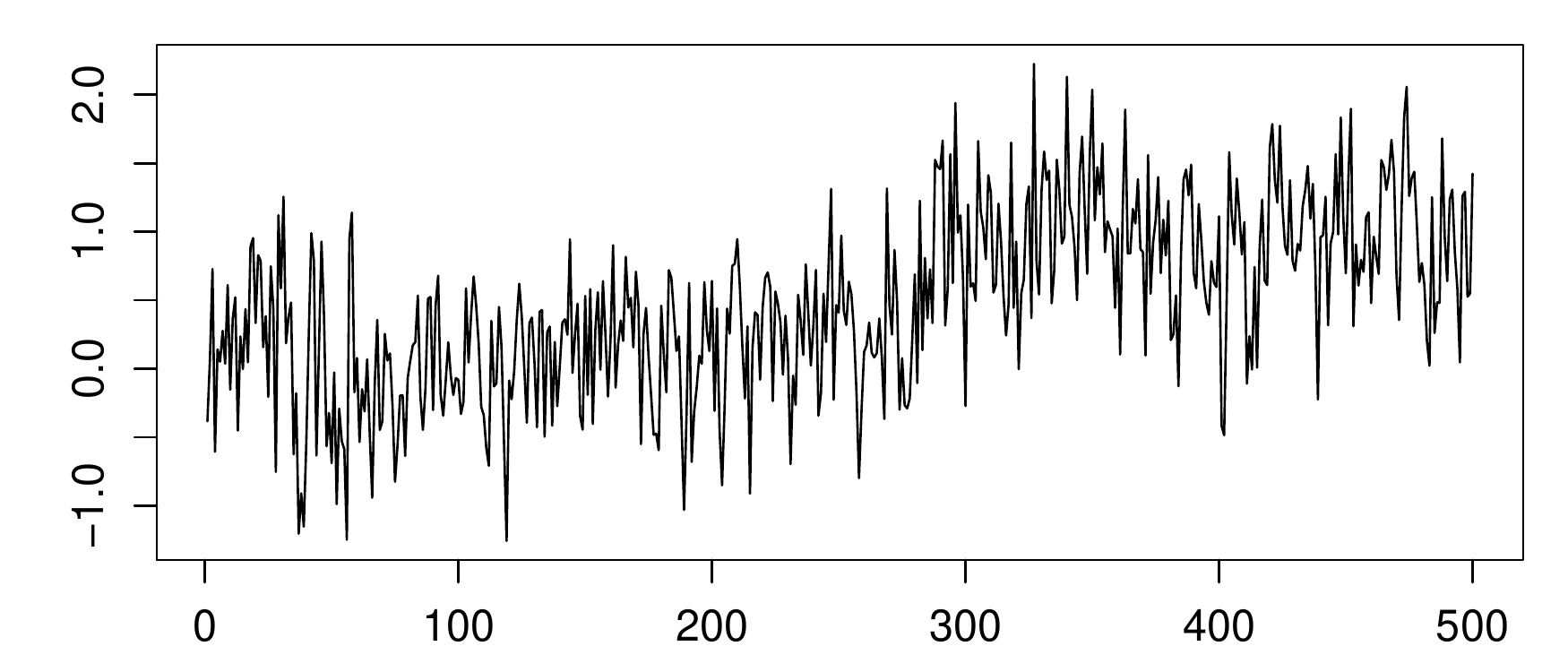} 
\caption{A typical sample path of length $T=500$ for model \eqref{mod1} with the mean function $\mu_2$.}\label{fig-data-mu}
\end{figure}

The results for the function $\mu_2$ are depicted in the lower part of Figure \ref{fig-mu}. The plots show that the upward bias is more pronounced than in the setting with $\mu_1$. This reflects the fact that it is more difficult to localize gradual  changes than a jump. In fact, it is quite hard to detect smooth time-variations on the interval $[0,u_0 + \delta]$ if $\delta$ is small. This is illustrated by Figure \ref{fig-data-mu}, which shows a typical sample path for 
model \eqref{mod1}  with mean function  $\mu_2$ of length $T=500$. As can be seen, the deviation of $\mu_2$ from zero is clearly visible only at time points which are somewhat larger than $u_0=0.5$. When getting close to $u_0$, the signal of time-variation becomes fairly weak and is more and more dominated by the noise of the error term.

In both designs, there is a certain fraction of estimates which take values below $u_0$. Theoretically, this fraction should be around $10\%$, since we have set the probability $\alpha$ of underestimating $u_0$ to equal $0.1$. In our simulations, however, the fraction obviously lies below the $10\%$ target as can be seen from the plots. This is a small sample effect which can be explained as follows: Our preliminary estimate $\hat{u}_0^{\circ}$ is quite conservative, tending to strongly overestimate $u_0$. Since $q_{\alpha}(u) \ge q_{\alpha}(u_0)$ for $u > u_0$, this implies that the estimate $\hat{\tau}_{\alpha} = q_{\alpha}(\hat{u}_0^{\circ})$ will often overshoot the value of the critical threshold $\tau_{\alpha} = q_{\alpha}(u_0)$, which is used to set up the second step estimator $\hat{u}_0$. As a result, the empirical probability of underestimating $u_0$ tends to lie below the target $\alpha$ in small samples.

\begin{figure}[H]
\centering
\includegraphics[width=0.32\textwidth]{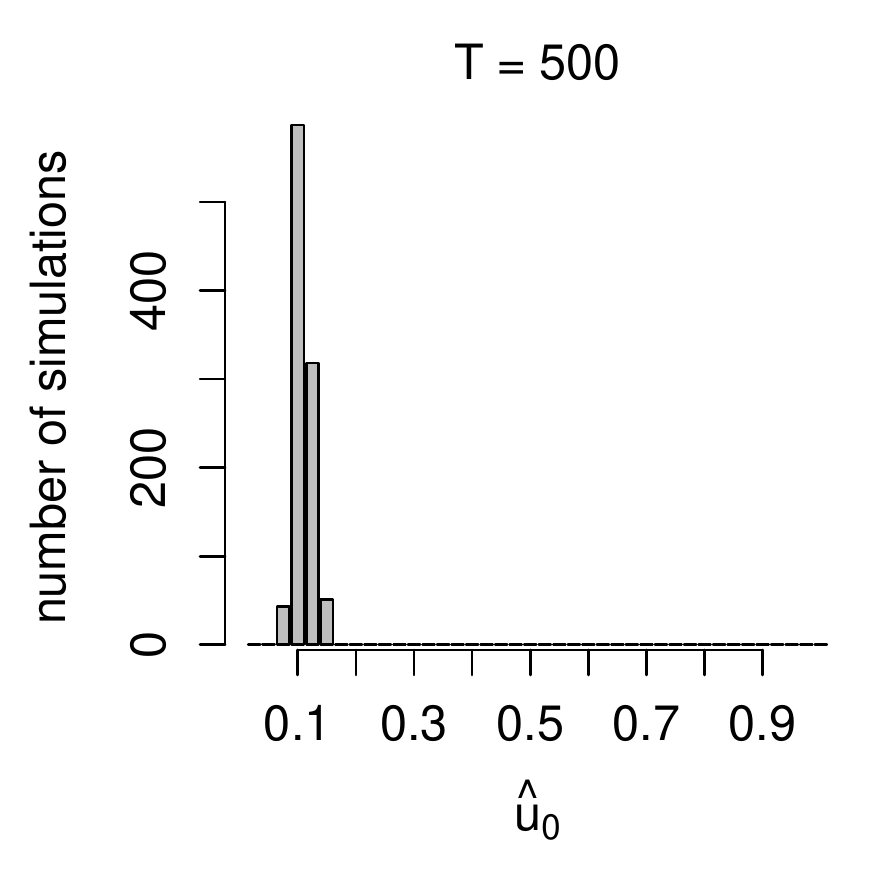}
\includegraphics[width=0.32\textwidth]{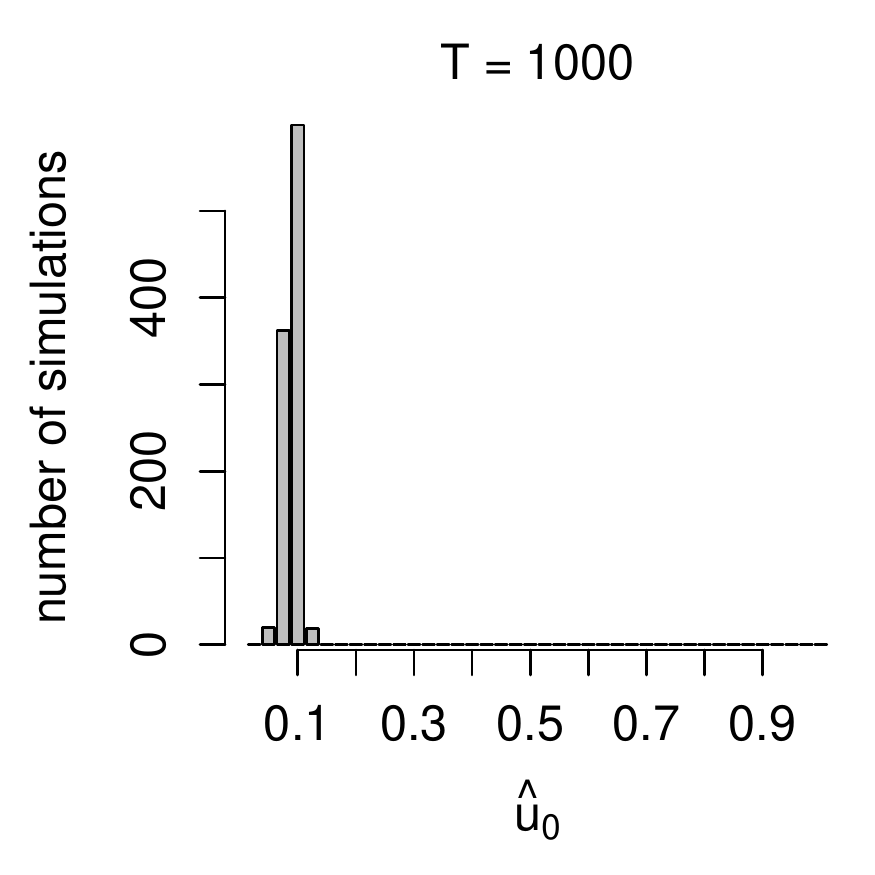}
\caption{Simulation results produced by our method in model (\ref{mod1}) with the mean function $\mu_3$ defined in \eqref{musmooth}.}\label{fig-bound0}
\end{figure}

We next investigate the performance of our procedure when the smooth change point $u_0$ occurs very early in the sample. In particular, we examine the extreme case where $u_0 = 0$ and the mean function is time-varying over the whole interval $[0,1]$. For this purpose, we consider the setting (\ref{mod1}) with the mean function
\begin{align}
\mu_3(u) & = 10 u \cdot 1( 0 \le u < 0.2) + \{2 - 2.5 (u - 0.2)\} \cdot 1 (u \ge 0.2). \label{musmooth}
\end{align}
The simulation results for this desgin are depicted in Figure \ref{fig-bound0} and show that our method detects the time-variation rather quickly. Of course, it is only able to detect it with some delay which becomes smaller when moving to the larger sample size $T = 1000$.

\subsection{Comparison with other methods}

In this section, we compare our estimation approach with the methods of \cite{Mallik2011, Mallik2013} and \cite{huskova1999} which are specifically designed to detect gradual changes in the location model \eqref{mod1}. As before, we assume that the mean function $\mu$ is constant on the time interval $[0,u_0]$, that is, $\mu(u) = \overline{\mu}$ for $u \le u_0$, and then starts to vary over time. The method of \cite{Mallik2011, Mallik2013}  allows to estimate the time point $u_0$ when $\mu$ is a smooth nonparametric function that is restricted to take values larger than $\overline{\mu}$ at time points $u > u_0$, that is, $\mu(u) > \overline{\mu}$ for $u > u_0$. The procedure of \cite{huskova1999} in contrast is based on the parametric assumption that $\mu(u) = \overline{\mu} + \delta \cdot (u - u_0)^{\beta} \cdot 1(u > u_0)$ for some slope parameter $\delta > 0$ and a known constant $\beta \in [0,1]$. In what follows, we set $\beta = 1$, thus considering Hu\u{s}kov\'{a}'s method for the class of broken lines with a kink at $u_0$.

To compare our method with these two approaches, we set $u_0 = 0.5$ and consider two different specifications of the mean function $\mu$,
\begin{align}
\mu_4(u) & = 2 (u - 0.5) \cdot 1(u > 0.5) \label{mu1} \\
\mu_5(u) & = \{10(u-0.5)\} \cdot 1( 0.5 < u < 0.6) + 1 (u \ge 0.6). \label{mu2}
\end{align}
Moreover, we let $\varepsilon_t$ be i.i.d.\ residuals that are normally distributed with mean zero and standard deviation $0.2$. Note that $\mu_4$ belongs to the parametric family of broken lines for which Hu\u{s}kov\'{a}'s method with $\beta = 1$ is designed. The function $\mu_5$, in contrast, is not an element of this parametric family.

Our estimator is implemented in the same way as in the simulation study of Subsection \ref{subsec-sim}. As the error terms are i.i.d., the error variance simplifies to $\sigma^2 = \ex[\varepsilon_t^2]$ and can be estimated as follows: Since $\mu$ is smooth, $\mu(\frac{t}{T}) - \mu(\frac{t-1}{T}) = O(T^{-1})$. This implies that $X_{t,T} - X_{t-1,T} = \varepsilon_t - \varepsilon_{t-1} + O(T^{-1})$, which in turn yields that $\ex(X_{t,T} - X_{t-1,T})^2 = \ex(\varepsilon_t - \varepsilon_{t-1})^2 + O(T^{-2}) = 2 \sigma^2 + O(T^{-2})$. Hence, we may simply estimate the error variance by $\hat{\sigma}^2 = T^{-1} \sum\nolimits_{t=2}^T (X_{t,T} - X_{t-1,T})^2/2$. This estimate is also used in the implementation of the method by \cite{Mallik2011, Mallik2013}. Hu\u{s}kov\'{a}'s estimator is constructed as described in equation (1.4) of \cite{huskova1999}. To implement the estimator of \cite{Mallik2011, Mallik2013}, we proceed as follows: Since the method is based on a Nadaraya-Watson smoother of $\mu$, we first select the bandwidth $h$ of this estimator. As shown in \cite{Mallik2013}, the rate-optimal bandwidth has the form $h = c T^{-1/(2k+1)}$, where $c$ is a constant and $\mu$ is assumed to have a cusp of order $k$ at the point $u_0$. This means that the first $(k-1)$ right derivatives of $\mu$ at $u_0$ are zero and the $k$-th right derivative is non-zero. For both functions, $\mu_4$ and $\mu_5$, $k$ is equal to $1$, implying that the optimal bandwidth is of the form $h = c T^{-1/3}$. Of course, since the order $k$ is unknown in practice, this is not a feasible choice of bandwidth. Moreover, even if $k$ were known, it is not clear how to pick the constant $c$. We here ignore these problems and pretend that $k$ is known. Having repeated the simulations for different choices of the constant $c$, we present the results for the choice $c = 0.1$ which yields the best  performance. For simplicity, we also assume that the baseline value $\overline{\mu}$ is known, so we do not have to replace it by an estimate.

\begin{figure}[H]
\centering
\includegraphics[width=0.32\textwidth]{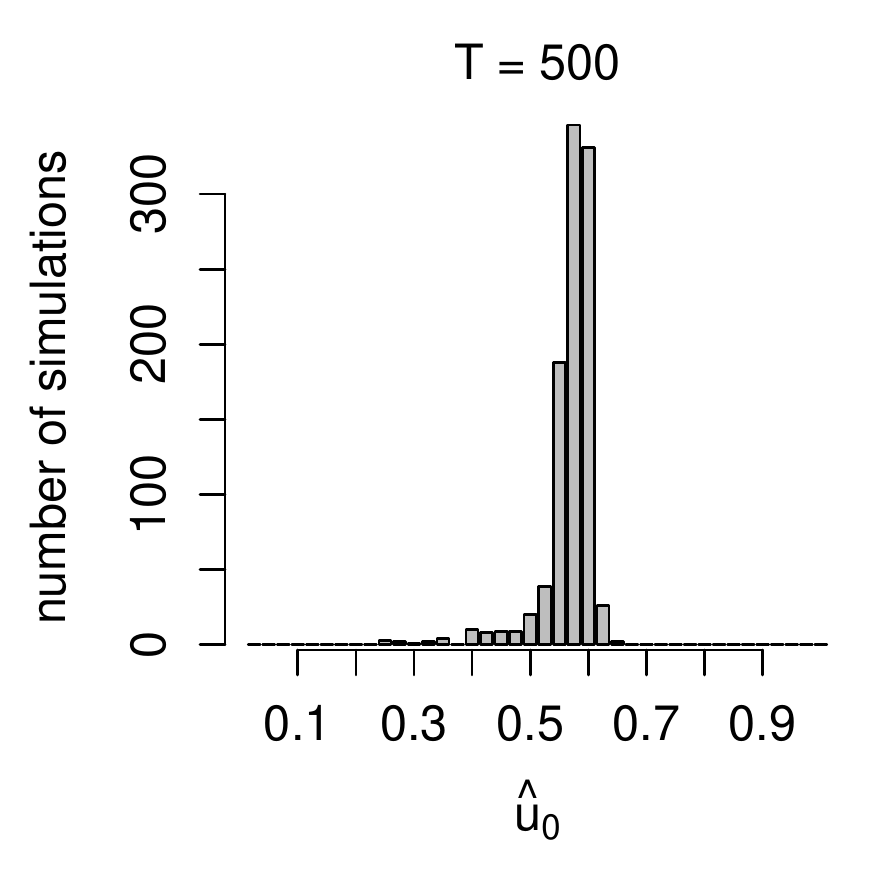}
\includegraphics[width=0.32\textwidth]{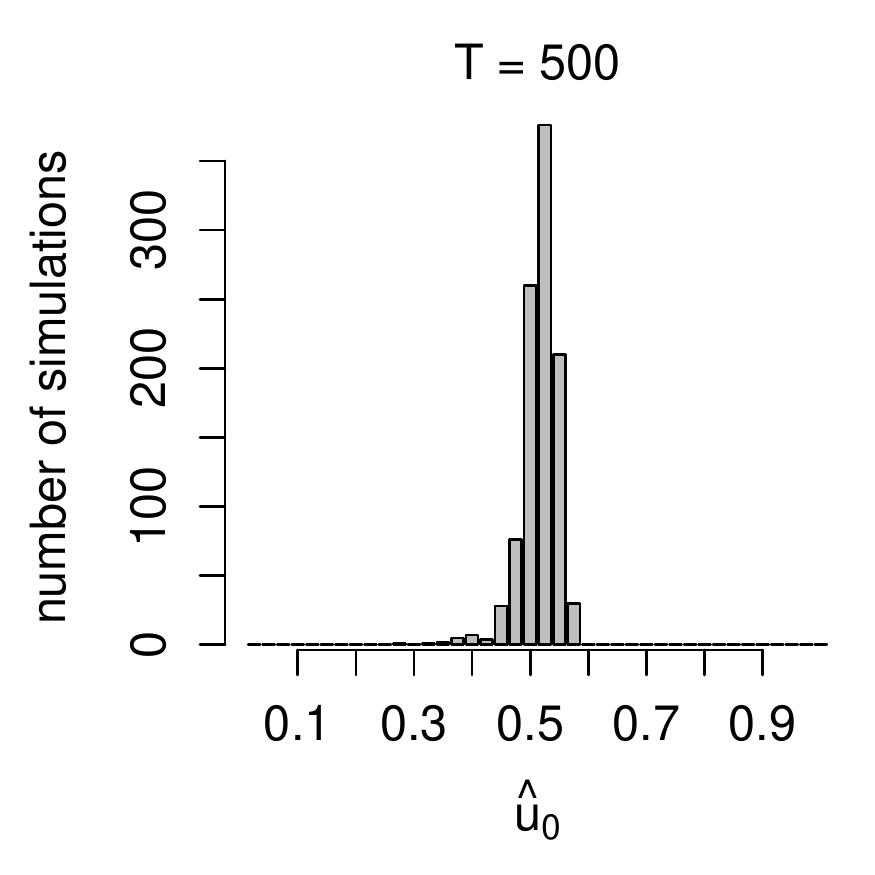}
\includegraphics[width=0.32\textwidth]{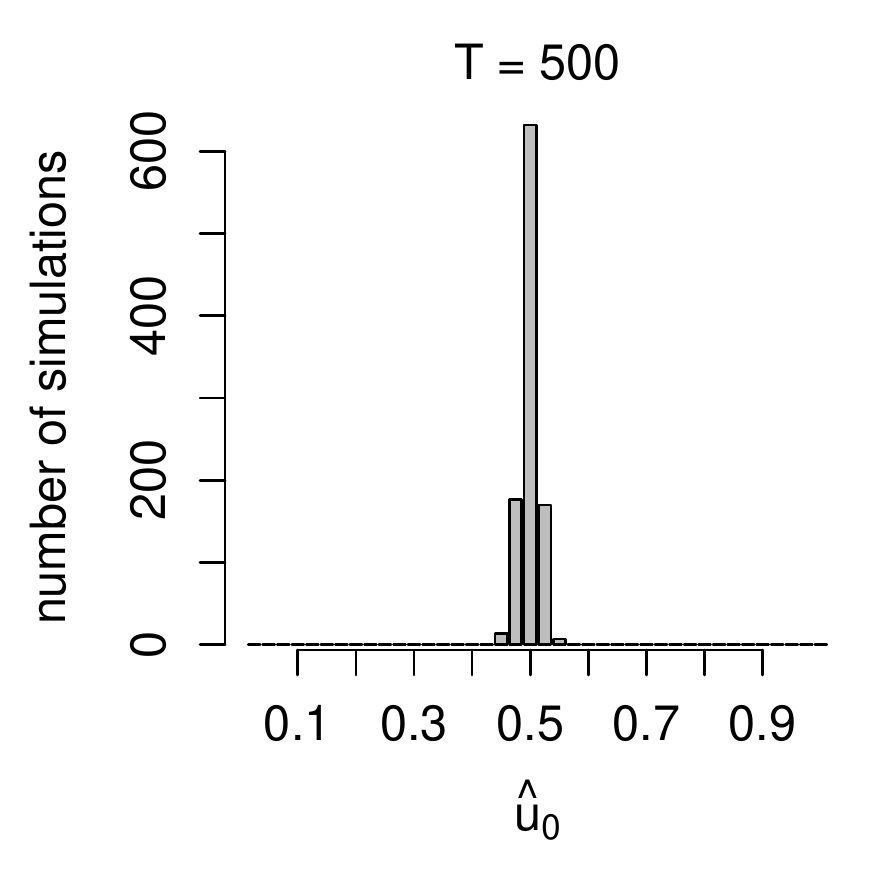} \\
\smallskip
\includegraphics[width=0.32\textwidth]{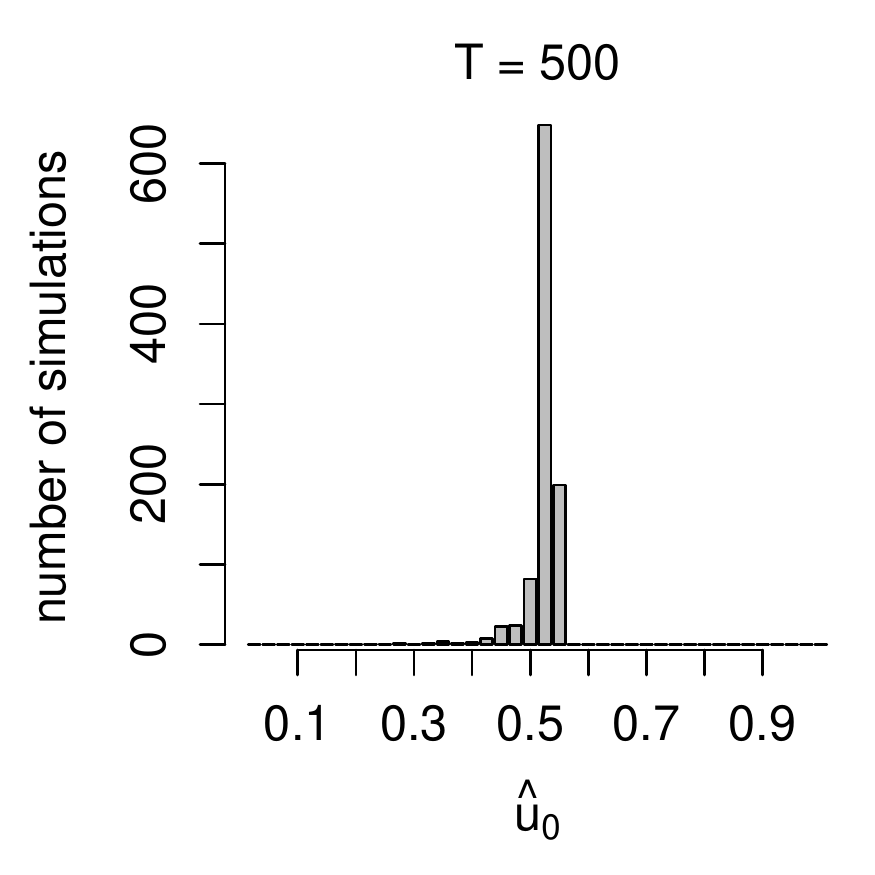}
\includegraphics[width=0.32\textwidth]{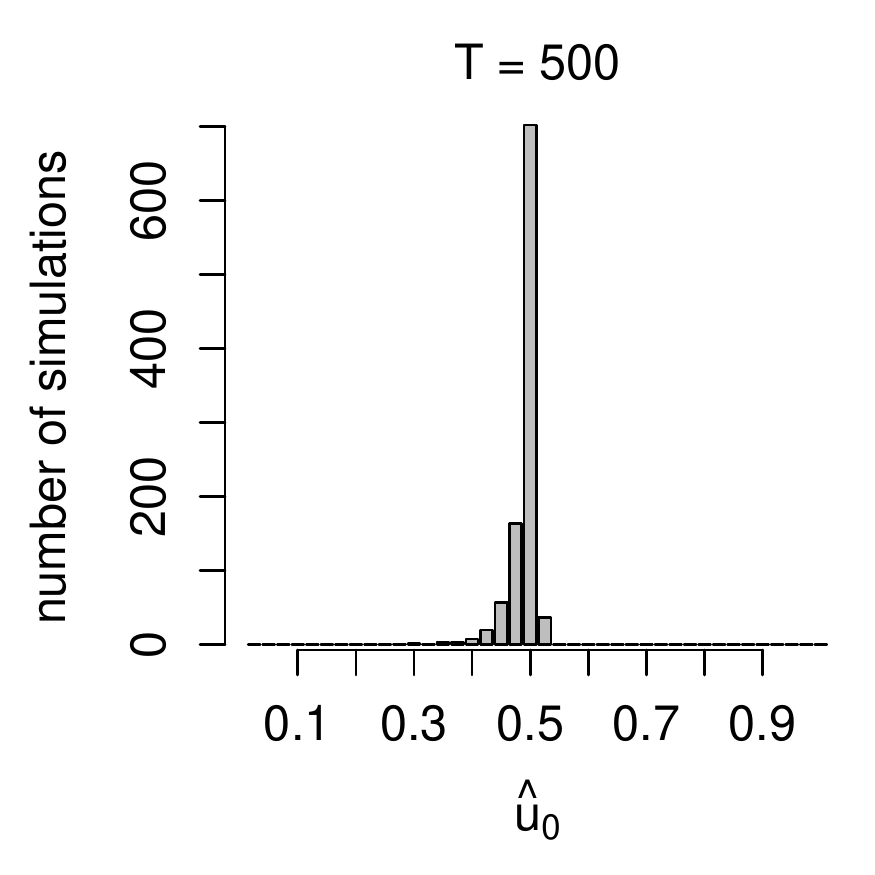}
\includegraphics[width=0.32\textwidth]{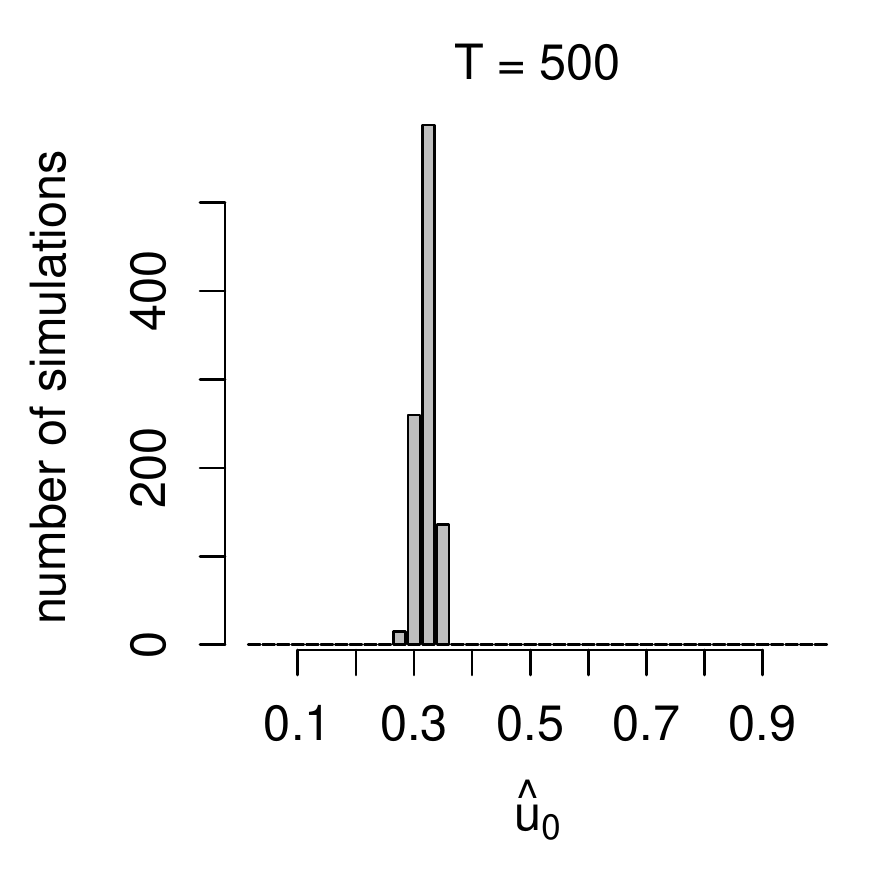}
\caption{Estimation results for model \eqref{mod1} with $\mu_4$ (upper panel) and $\mu_5$ (lower panel). The left-hand plots correspond to our method, the middle ones to the approach of \cite{Mallik2011, Mallik2013} and the right-hand ones to the procedure in \cite{huskova1999}.}\label{fig-compare-1}
\end{figure}

The results for the regression function $\mu_4$ are presented in the upper part of Figure \ref{fig-compare-1}. As can be seen, Hu\u{s}kov\'{a}'s method outperforms both ours and the $p$-value based approach of \cite{Mallik2011, Mallik2013}. This is not surprising since it is tailored to a specific parametric class of mean functions to which $\mu_4$ belongs. Even though less precise than Hu\u{s}kov\'{a}'s estimator, both our and the $p$-value based method perform well, ours tending to be a bit more upward biased and thus slightly more conservative. The results for the regression function $\mu_5$ are presented in the lower part of Figure \ref{fig-compare-1}. As before, both our method and that of Mallik et al.\ perform quite well. The parametric method of \cite{huskova1999}, in contrast, completely fails to provide reasonable estimates of $u_0$. The reason for this is simply that $\mu_5$ does not satisfy the parametric assumptions of this approach.

\begin{figure}[H]
\centering
\includegraphics[width=0.32\textwidth]{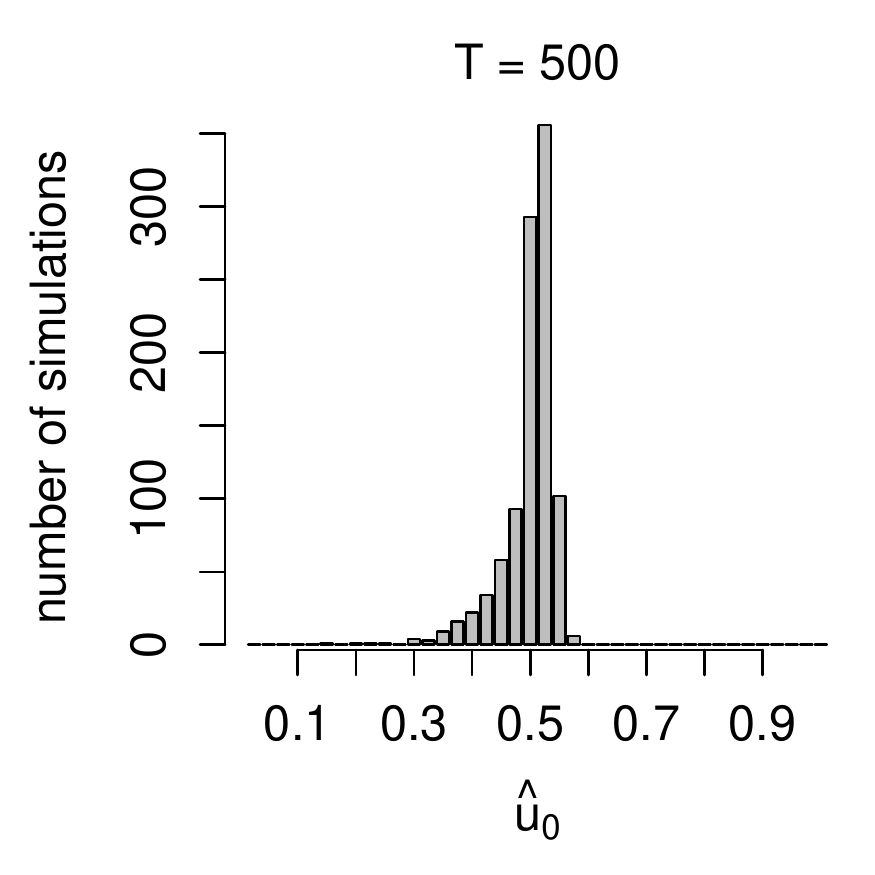}
\includegraphics[width=0.32\textwidth]{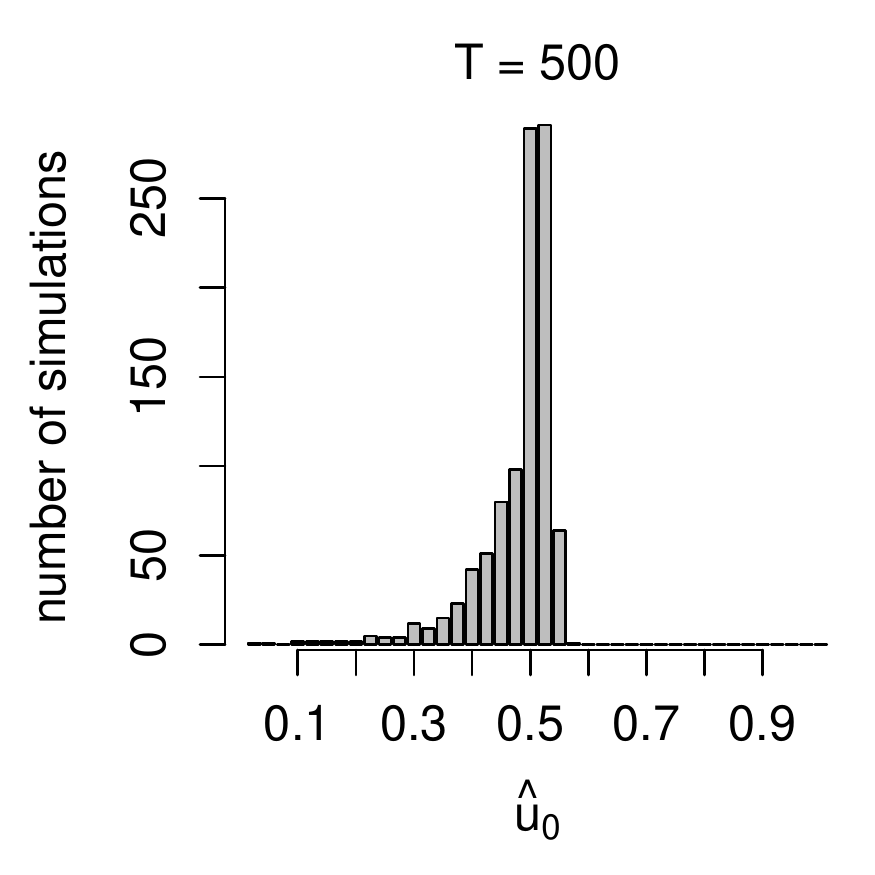}
\includegraphics[width=0.32\textwidth]{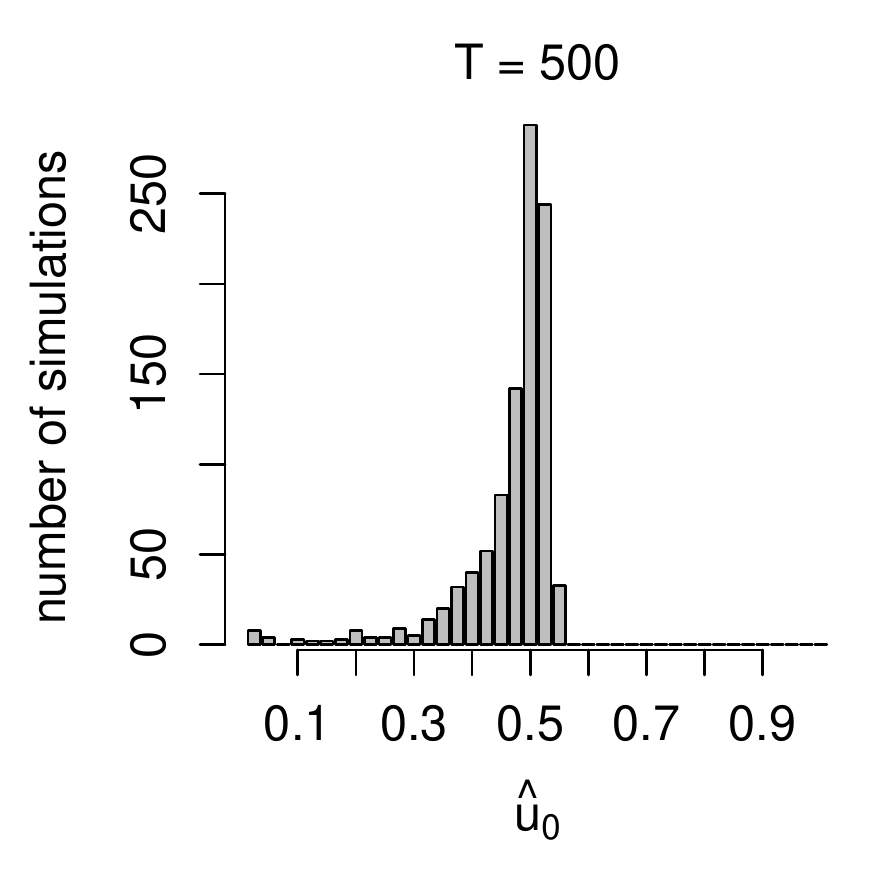}
\caption{Results for the method of \cite{Mallik2011, Mallik2013} in model \eqref{mod1} with $\mu_4$ and the bandwidth $h = c T^{-1/3}$, where $c = 0.2$ (left), $c = 0.3$ (middle) and $c = 0.4$ (right).}\label{fig-compare-2}
\end{figure}

To implement the method of \cite{Mallik2011, Mallik2013}, we have used an optimally tuned bandwidth which presupposes knowledge of the degree of smoothness $k$ and have treated the mean value $\overline{\mu}$ as known. Nevertheless, this approach only provides slightly better results than ours. In practice, $\overline{\mu}$ must of course be estimated and the optimal choice of bandwidth is not available. Moreover, the performance of the method varies quite considerably with the bandwidth. This is illustrated in Figure \ref{fig-compare-2} which shows the estimation results when picking the bandwidth to be rate optimal with the constants $c = 0.2, 0.3, 0.4$ instead of $c=0.1$.\footnote{Note that for all these values of $c$, the bandwidth is fairly small, resulting in an undersmoothed estimate of the mean function $\mu$. Specifically, for a sample size of $T=500$, the choice $c = 0.1$ corresponds to a bandwidth window of approximately $5$ data points and $c = 0.4$ to a window of $25$ points. Indeed, the method appears only to work in a reasonable way when strongly undersmoothing, which is already indicated by the fact that the optimal bandwidth is of the rate $T^{-1/3}$.} As can be seen, the results get much worse when slightly changing the bandwidth parameter $c$, a large fraction of the estimates tending to strongly underestimate $u_0$.

The above discussion points to an important advantage of our method: The tuning parameter $\tau_\alpha$ on which it depends is much more harmless than a bandwidth parameter. As $\alpha$ can be interpreted in terms of the probability of underestimating $u_0$, it is clear how to choose $\tau_\alpha$ in a reasonable way in practice. Hence, we do not run the risk of producing poor estimates by picking the tuning parameter in an inappropriate way. This makes our procedure particularly attractive to apply in practice. We finally point out that the new method is not specifically designed for detecting a change in the nonparametric location model \eqref{mod1} but can be easily adapted to other change point problems. This is illustrated in the Supplementary Material, where we show results for a nonparametric volatility model.

\subsection{Applications}

We now apply the proposed  estimation method to the data presented  in the Introduction. We first consider the monthly temperature anomalies of the northern hemisphere from 1850 to 2013 depicted in the left-hand panel of Figure \ref{fig1}. These anomalies are temperature deviations from the average 1961--1990 measured in degrees Celsius. The data set is called HadCRUT4 and can be obtained from the Climatic Research Unit of the University of East Anglia, England. A detailed description of the data can be found in \cite{Brohan2006}.

Inspecting the temperature data, they can be seen to exhibit a seasonal as well as a trending behaviour. We thus model them by the equation
\begin{equation}\label{model1-app}
X_{t,T} = s(t) + \mu \Bigl(\frac{t}{T}\Bigr) + \varepsilon_{t,T} \qquad (t=1,\ldots,T), 
\end{equation}
where $s$ is a seasonal component with a period of $12$ months, $\mu$ is a nonparametric trend and $\varepsilon_{t,T}$ are error terms with zero mean. For identification, we assume that $\sum\nolimits_{t=1}^{12} s(t) = 0$. From the data plot, one can also see a larger variance at the beginning of the sample, suggesting that the errors are nonstationary. To pick up these nonstationary effects, we allow the error terms $\varepsilon_{t,T}$ to be locally stationary. Rescaling the time argument of the trend component while letting the periodic component depend on real time is a rather natural way to formulate the model. It captures the fact that the trend function is much smoother and varies more slowly than the seasonal part. Analogous model formulations can be found for example in \cite{SubbaRao2004} and  \cite{Atak2011}.

The issue of global warming has received much attention over the last decades. One question of interest is to locate the onset of the warming trend; see e.g.\ \cite{Thanasis2011} or \cite{Mallik2011} for a statistical analysis of this question. The challenge is thus to estimate the time point $u_0$ where the function $\mu$ starts to strongly trend upwards. To clarify this issue, we apply our estimation method to the anomaly data at hand. Importantly, we do not have to pre-process the data and deseasonalize them but can work with the raw data itself. The reason for this is as follows: The seasonal component $s$ shows up in averages of the form $A_T(w) = T^{-1} \sum\nolimits_{t=1}^{\lfloor wT \rfloor} s(t)$ in the statistic $\hat{\Dsup}_T$ which underlies our estimation procedure. Since $\sum\nolimits_{t=1}^{12} s(t) = 0$ by our normalization, it holds that $A_T(w) = O(T^{-1})$ uniformly in $w$. Hence, the seasonal component gets smoothed or averaged out when calculating the statistic $\hat{\Dsup}_T$, implying that we can simply ignore it.

To implement our procedure, we proceed as described in Section \ref{sec-impl} and set $\alpha = 0.1$. To calculate the quantiles in Step 1 of the implementation, we use an estimator of the form \eqref{est-lrv} with a bandwidth $h$ that corresponds to approximately $10$ years of data and the bandwidth $b = 15$, meaning that we take into account the first $15$ autocovariances when computing the HAC estimator. With these choices, we obtain an estimate $\hat{u}_0$ which corresponds to the year $1915$ and is graphically illustrated by the dashed vertical line in the left-hand panel of Figure \ref{fig1}. As a robustness check, we have varied the bandwidth $h$ between $5$ and $15$ years and $b$ between $10$ and $20$. For all these choices, we obtain estimates roughly between $1910$ and $1920$, providing evidence that the mean temperature starts to trend upwards in this time region. This finding is in broad accordance with other analyses. \cite{Woodroofe2012} for example apply isotonic regression techniques to the data set of yearly global temperature anomalies and find that the warming trend emerges around the same time.

We next turn to the daily return data of the S\&P 500 index which are depicted in the right-hand panel of Figure \ref{fig1}. A simple locally stationary model for financial returns is given by the equation
\begin{equation}\label{return-model}
r_{t,T} = \sigma \Bigl(\frac{t}{T}\Bigr) \varepsilon_t,
\end{equation}
where $r_{t,T}$ denotes the daily return, $\sigma$ is a time-varying volatility function and $\varepsilon_t$ are i.i.d.\ residuals with zero mean and unit variance. Model \eqref{return-model} has been studied in a variety of papers [see 
\cite{Drees2003} and  \cite{Fryzlewicz2006} among others]. In many situations, it is realistic to assume that the volatility level is more or less constant within some time span $[u_0,1]$, where $u=1$ is the present time point, and remains roughly constant in the near future $(1,1+\delta]$. In this case, $\sigma(u) \approx \sigma(1)$ at future time points $u \in (1,1+\delta]$, which suggests to use the present volatility level $\sigma(1)$ as a forecast for the near future [see \cite{Fryzlewicz2006} among others]. To obtain a good volatility forecast, we thus have to construct a good estimator of $\sigma(1)$. If we knew the time point $u_0$, we could come up with a very simple and precise estimator. In particular, we could estimate $\sigma^2(1)$ by the sample variance of the observations contained in the time interval $[u_0,1]$. In practice, however, the time point $u_0$ is not observed and has to be estimated.

In what follows, we estimate the time span $[u_0,1]$ where the volatility level of the S\&P 500 returns from Figure \ref{fig1} is more or less constant. To do so, we have to reformulate our estimation method, since it is designed to apply to time spans of the form $[0,u_0]$ rather than $[u_0,1]$. Since this is trivial to achieve and simply a matter of notation, we neglect the details. As time-variation in the volatility is equivalent to time-variation in the variance $\var(r_{t,T}) = \ex[r_{t,T}^2]$, we set up our procedure to detect changes in the variance and implement it as described in Section \ref{sec-impl}. As before, we let $\alpha = 0.1$. Moreover, we choose $h = 0.1$, noting again that the results are very robust to different choices of $h$. Finally, we set the bandwidth $b$ to equal zero, assuming that the return data are independent. Our estimate $\hat{u}_0$ of the time point $u_0$ is depicted as the vertical dashed line in the right-hand panel of Figure \ref{fig1}.

\bigskip
\textbf{Acknowledgements.}
This work has been supported in part by the Collaborative Research Center ``Statistical modeling of nonlinear dynamic processes'' (SFB 823, Teilprojekt A1, C1) of the German Research Foundation. We would like to thank Rainer Dahlhaus, Wolfgang Polonik and Stanislav Volgushev for helpful discussions and comments on an earlier version of this manuscript. We are also grateful to Alina Dette and Martina Stein, who typed parts of this paper with considerable technical expertise. The constructive comments of an associate editor and two referees on an earlier version of this paper led to a substantial improvement of the manuscript. Parts of this paper were written  while the authors were visiting the Isaac Newton Institute, Cambridge, UK, in 2014 (``Inference for change-point and related processes'') and the authors would like to thank the institute for its hospitality.

\appendix
\section*{Appendix}
\def\theequation{A.\arabic{equation}}
\setcounter{equation}{0}

In this appendix, we prove the main theoretical results of the paper. Throughout the appendix, the symbol $C$ denotes a generic constant which may take a different value on each occurrence. Moreover, the expression $\| X \|_p = (\ex |X|^p)^{1/p}$ is used to denote the $L_p$-norm of a real-valued random variable $X$.

\subsection*{Auxiliary Results}

Before we turn to the proofs of the main theorems, we derive some technical lemmas which are needed later on. To formulate them, we introduce some additional notation. To start with, partition the observations $\{X_{t,T}: t=1,\ldots,T\}$ into blocks of size $q$, where the $r$-th block spans the observations from time point $(r-1)q+1$ to $rq$ and we set $q = CT^b$ for some small $b > 0$ (in particular $b < \frac{1}{4}$). Now define 
\[ W_T(k,k^\prime) = \sup_{f \in \mathcal{F}} \Bigl| \sum\limits_{r=k}^{k^\prime} Q_{r,T}(f) \Bigr| \]
along with
\[ Q_{r,T}(f) = \frac{1}{\sqrt{(k^\prime - k + 1)q}} \sum\limits_{t=(2r-2)q+1}^{(2r-1)q \wedge T} \bigl( f(X_{t,T}) - \ex f(X_{t,T}) \bigr). \]
The terms $Q_{r,T}(f)$ are scaled sums of the variables $f(X_{t,T}) - \ex f(X_{t,T})$, the summation running over the observations of the $(2r-1)$-th block. The expression $W_T(k,k^\prime)$ sums up the terms $Q_{k,T}(f),\ldots,Q_{k^\prime,T}(f)$ which correspond to the odd blocks $(2k-1),(2k+1),(2k+3),\ldots,(2k^\prime-1)$. The next two lemmas provide a bound on the $L_p$-norm of $ W_T(k,k^\prime)$.

\begin{lemmaA}\label{lemmaA1}
Let assumptions (C\ref{A1}) and (C\ref{A2}) be satisfied and let $f_0 \in \mathcal{F}$ have the property that $\ex |f_0(X_{t,T})|^{(1+\delta)p} \le C$ for some even $p \in \naturals$ and a small $\delta > 0$. Then 
\[ \Bigl\| \sum\limits_{r=k}^{k^\prime} Q_{r,T}(f_0) \Bigr\|_p \le C \] 
for some sufficiently large constant $C$.
\end{lemmaA}

\textbf{Proof.} To shorten notation, write $w_{t,T} = f_0(X_{t,T}) - \ex f_0(X_{t,T})$ and consider the term
\begin{align*}
V_T 
 & = V_T(k,k^\prime) = \ex \Bigr[ \Bigl( \sum\limits_{r=k}^{k^\prime} Q_{r,T}(f_0) \Bigr)^p \Bigr] \\
 & \leq \frac{1}{((k^\prime - k + 1)q)^{p/2}} \sum_{r_1,\dots,r_p =k}^{k^\prime} \sum_{t_1=(2r_1-2)q+1}^{(2r_1-1)q \wedge T} 
 \ldots \sum_{t_p=(2r_p-2)q + 1}^{(2r_p-1)q \wedge T} \bigl| \ex [w_{t_1,T} \dots w_{t_p,T}] \bigr| \nonumber \\
 &\leq \frac{p!}{((k^\prime - k + 1)q)^{p/2}} \sum_{{t_1,\dots,t_p = (2k-2)q+1 \atop t_1 \leq \ldots \leq t_p}}^{(2k^\prime-1)q \wedge T} \bigl| \ex [w_{t_1,T} \dots w_{t_p,T}] \bigr|. 
\end{align*}

Let $(t_1,\ldots,t_p)$ be a tuple of ordered indices, that is, $t_1 \le \ldots \le t_p$. We say that the index $t_i$ has a neighbour if $|t_i - t_{i-1}| \le C^* \log T$ or $|t_i - t_{i+1}| \le C^* \log T$ for some large constant $C^*$ to be specified later on. Moreover, $t_i$ is said to have exactly one neighbour if either $|t_i - t_{i-1}| \le C^* \log T$ and $|t_i - t_{i+1}| > C^* \log T$ or vice versa. Finally, we call $(t_{i-1},t_i)$ a pair of neighbours if $|t_i - t_{i-1}| \le C^* \log T$. Now let $S_{\leq}$ denote the set of ordered tuples $(t_1,\dots,t_p) \in \{ (2k-2)q+1, \dots, (2k^\prime - 1) q \wedge T \}^p$ such that each index $t_i$ has a neighbour. In addition, let $S_>$ be the set of tuples such that at least one index does not have a neighbour. With this notation at hand, we can write $V_T = V_T^{\leq} + V_T^{>}$, where for $\ell \in \{ \leq, > \}$,
\[ V_T^{\ell} = \frac {p!}{((k^\prime - k + 1)q)^{p/2}} \sum_{(t_1,\dots,t_p) \in S_{\ell}} \bigl| \ex[ w_{t_1,T} \dots w_{t_p,T} ] \bigr|. \]

We now analyze the two terms $V_T^{\leq}$ and $V_T^{>}$ separately. For the investigation of $V_T^{\leq}$, define
\[ S_{\leq, a} = \bigl\{ (t_1,\dots,t_p) \in S_{\leq} \ | \ \mbox{each index $t_i$ has exactly one neighbour} \bigr\} \]
together with $S_{\leq,b} = S_{\leq} \setminus S_{\leq,a}$. First suppose that $(t_1,\dots,t_p) \in S_{\leq,a}$. In this case, there are exactly $p/2$ pairs $(t_{2i-1},t_{2i})$ of neighbours (recalling that $p$ is even by assumption). Using Davydov's inequality (see e.g.\ Corollary 1.1 in \cite{Bosq1996}) to bound the covariances of the mixing variables $w_{t,T}$, we obtain that
\begin{align*}
\bigl| \ex [w_{t_1,T} \dots w_{t_p,T}] \bigr| 
 & \leq  \bigl| \ex [w_{t_1,T} w_{t_2,T}] \ex [w_{t_3,T} \dots w_{t_p,T}] \bigr| + \bigl| \cov (w_{t_1,T} w_{t_2,T}, w_{t_3,T} \dots w_{t_p,T}) \bigr| \\
 & = \bigl| \ex [w_{t_1,T} w_{t_2,T}] \ex [w_{t_3,T} \dots w_{t_p,T}] \bigr| + O \bigl( \alpha (C^* \log T ) \bigr) \\
 & = \bigl| \cov (w_{t_1,T}, w_{t_2,T}) \ex [w_{t_3,T} \dots w_{t_p,T}] \bigr| + O \bigl( \alpha (C^* \log T ) \bigr) \\
 & \ \ \vdots \\
 & \leq \Bigl| \prod_{i=1}^{p/2} \cov (w_{t_{2i-1},T}, w_{t_{2i},T}) \Bigr| + O (T^{-\nu }),
\end{align*}
where we have used the fact that the mixing coefficients are decaying exponentially fast and the constant $\nu > 0$ can be made arbitrarily large (by choosing the constant $C^*$ sufficiently large). This implies that
\begin{align*}
V_T^{\leq,a}
 & = \frac{p!}{((k^\prime - k + 1)q)^{p/2}} \sum_{(t_1,\dots,t_p) \in S_{\leq,a}} \bigl| \ex [ w_{t_1,T} \dots w_{t_p,T} ] \bigr| \\
 & \leq \frac{p!}{((k^\prime - k + 1)q)^{p/2}} \sum_{(t_1,\dots,t_p) \in S_{\leq,a}} \Bigl| \prod_{i=1}^{p/2} \cov (w_{t_{2i-1},T}, w_{t_{2i},T}) \Bigr| + o(1) \\
 & \leq \frac{p!}{((k^\prime - k + 1)q)^{p/2}} \prod_{i=1}^{p/2} \Bigr( \sum_{\ell = 0}^{\lceil C^* \log T \rceil} \sum_{t_{2i-1}=(2k-2)q+1}^{(2k^\prime-1) q \wedge T} \bigl| \cov(w_{t_{2i-1},T}, w_{t_{2i-1}+\ell,T}) \bigr| \Bigl) + o(1) \\
 & \leq C \frac{p!}{((k^\prime - k + 1)q)^{p/2}} ((k^\prime - k + 1)q)^{p/2} \Bigl ( \sum_{\ell = 0}^{\lceil C^* \log T \rceil} \alpha(\ell) \Bigr)^{p/2} + o(1) \leq C
\end{align*}
for some sufficiently large constant $C$, where the last line again uses Davydov's inequality to bound the covariance expressions in the formula.

Next consider the sum $V_T^{\leq,b}$ corresponding to indices in the set $S_{\leq,b}$. The cardinality of this set is bounded by $C (( k^\prime - k + 1)q)^{\frac{p}{2}-1} (\log T)^{\frac{p}{2}+1}$, which implies
\begin{align*}
V_T^{\leq,b} & = \frac {p!}{((k^\prime - k + 1)q)^{p/2}} \sum_{(t_1,\dots,t_p) \in S_{\leq,b}} \bigl| \ex [w_{t_1,T} \dots w_{t_p,T} ] \bigr| \leq C \frac {(\log T)^{p/2+1}}{(k^\prime - k + 1)q} = o(1) 
\end{align*}
(noting that $q = T^b$). This shows that the term $V_T^{\leq}$ is bounded.

Finally, we examine the term $V_T^{>}$ corresponding to the index set $S_>$. By definition, the tuples contained in this set have at least one element, say $t_i$, without a neighbour, that is, $|t_i-t_{i+1}| > C^* \log T$ and $|t_i-t_{i-1}| > C^* \log T$. Exploiting the mixing conditions on the model variables in a similar way as above, we obtain that
\begin{align*}
\ex [ w_{t_1,T}\dots w_{t_p,T}] 
 & = \ex [w_{t_1,T} \dots w_{t_{i-1},T}] \ex [w_{t_i,T} \dots w_{t_p,T}] + \cov(w_{t_1,T} \dots w_{t_{i-1},T}, w_{t_i,T} \dots w_{t_p,T}) \\
 & = \ex [w_{t_1,T} \dots w_{t_{i-1},T}] \cov(w_{t_i,T}, w_{t_{i+1},T} \dots w_{t_p,T}) + O(T^{- \nu})   
   =  O(T^{- \nu}),
\end{align*}
where $\nu$ can be chosen arbitrarily large (if $C^*$ is chosen large enough). Recalling the definition of $V_T^{>}$, this yields that $V_T^{>} = o(1)$. Putting everything together, the quantity $V_T$ is seen to be bounded. This completes the proof. \qed

\begin{lemmaA}\label{lemmaA2}
Let (C\ref{A1}) and (C\ref{A2}) be satisfied. Moreover, assume that for some even $p \in \naturals$ and some small $\delta > 0$,
\[ \ex \Bigl[ \Bigl| \frac{f(X_{t,T}) - f^\prime(X_{t,T})}{\dist(f,f^\prime)} \Bigr|^{(1 + \delta) p} \Bigr] \le C \]
for all functions $f, f^\prime \in \mathcal{F}$. Then for any $f_0 \in \mathcal{F}$, 
\begin{equation*}
\bigl\| W_T(k,k^\prime) \bigr\|_p \le C \Bigl( \Bigl\| \sum\limits_{r=k}^{k^\prime} Q_{r,T}(f_0) \Bigr\|_p + \int_0^{\textnormal{diam}(\mathcal{F})} \mathcal{N}(w/2,\mathcal{F},\dist)^{1/p} dw \Bigr), 
\end{equation*}
where $\mathcal{N}(w,\mathcal{F},\dist)$ is the covering number of $(\mathcal{F},\dist)$ and $\textnormal{diam}(\mathcal{F}) = \sup_{f,f^{\prime} \in \mathcal{F}} \dist(f,f^{\prime})$ denotes the diameter of $\mathcal{F}$. 
\end{lemmaA}

\textbf{Proof.} The claim immediately follows from Theorem 2.2.4 and Corollary 2.2.5 in \cite{vanderVaart1996} (see their remark on p.100 before Subsection 2.2.1). It thus suffices to verify the conditions of Theorem 2.2.4. In particular, we have to show that
\begin{equation*}
\ex \Bigl[ \Bigl| \sum_{r=k}^{k^\prime} Q_{r,T}(f) - \sum_{r=k}^{k^\prime} Q_{r,T}(f^\prime) \Bigr|^p \Bigr] \leq C \dist(f,f^\prime)^p
\end{equation*}
for some sufficiently large constant $C$. To prove this, we introduce the notation
\[ w_{t,T} = \frac{f(X_{t,T}) - f^\prime(X_{t,T})}{\dist(f,f^\prime)} - \ex \Bigl[ \frac{f(X_{t,T})- f^\prime(X_{t,T})}{\dist(f,f^\prime)} \Bigr] \]
and consider
\begin{align*}
V_T 
 & = V_T (k,k^\prime) = \ex \Bigr[ \Bigl| \sum_{r=k}^{k^\prime} \frac{Q_{r,T}(f) - Q_{r,T}(f^\prime)}{\dist(f,f^\prime)} \Bigr|^p \Bigr] \nonumber \\*
 & \leq \frac{1}{((k^\prime - k + 1)q)^{p/2}} \sum_{r_1,\dots,r_p =k}^{k^\prime} \sum_{t_1=(2r_1-2)q+1}^{(2r_1-1)q \wedge T} \ldots \sum_{t_p=(2r_p-2)q+1}^{(2r_p-1)q \wedge T} \bigl| \ex [w_{t_1,T} \dots w_{t_p,T}] \bigr| \\
 &\leq \frac{p!}{((k^\prime - k + 1)q)^{p/2}} \sum_{{t_1,\dots,t_p = (2k-2)q+1 \atop t_1 \leq \dots \leq t_p}}^{(2k^\prime-1)q \wedge T} \bigl| \ex [w_{t_1,T} \dots w_{t_p,T}] \bigr|. 
\end{align*}
Repeating the arguments from Lemma \ref{lemmaA1}, we can show that $V_T$ is bounded, thus completing the proof. \qed

\subsection*{Proof of Theorem \ref{theo-measure}}

To show that $\hat{\Hpt}_T = \sqrt{T} [\hat{\Dpt}_T - \Dpt]$ weakly converges to $\Hpt$, it suffices to prove that
\begin{equation}\label{theo1-1}
\hat{\Hpt}_T^c := \sqrt{T} \bigl[ \hat{\Dpt}_T - \ex \hat{\Dpt}_T \bigr] \convw \Hpt  
\end{equation}
together with
\begin{equation}\label{theo1-2}
\sqrt{T} \sup_{(u,v,f) \in \Delta \times \mathcal{F}} | \ex \hat{\Dpt}_T - \Dpt | = o(1),
\end{equation}
where $\hat{\Hpt}_T^c$ is the centred version of $\hat{\Hpt}_T$.
We start with the proof of \eqref{theo1-2}. Making use of condition (C\ref{A4}), we obtain that 
\begin{align*}
\frac{1}{\sqrt{T}} \sum\limits_{t = 1}^{\lfloor uT \rfloor} \ex \bigl[f(X_{t,T})\bigr]
 & = \frac{1}{\sqrt{T}} \sum\limits_{t = 1}^{\lfloor uT \rfloor} \ex \Bigr[f\Big(X_t\Big(\frac{t}{T}\Big)\Big)\Bigr] + o(1) \\
 & = \sqrt{T} \sum\limits_{t = 1}^{\lfloor uT \rfloor} \int_{\frac{t-1}{T}}^{\frac{t}{T}} \ex \bigl[f(X_t(w))\bigr] dw + o(1) \\
 & = \sqrt{T} \int_0^u \ex \bigl[f(X_t(w))\bigr] dw + o(1)
\end{align*}
uniformly with respect to $u \in [0,1]$ and $f \in {\cal F}$. From this, \eqref{theo1-2} immediately follows. 
To verify \eqref{theo1-1}, we show weak convergence of the finite dimensional distributions of $\hat{\Hpt}_T^c$ as well as stochastic equicontinuity of $\hat{\Hpt}_T^c$. In particular, we derive the following two results. 
\begin{propositionA}\label{propA1}
For any finite number of points $(u_i,v_i,f_i)$ with $1 \le i \le n$, it holds that
\[ ( \hat{\Hpt}_T^c(u_1,v_1,f_1),\ldots,\hat{\Hpt}_T^c(u_n,v_n,f_n))^{^\intercal} \convd \normal(0,\Sigma) \]
where $\Sigma = (\Sigma_{ij})_{1 \le i,j \le n}$ and $\Sigma_{ij} = \cov(\Hpt(u_i,v_i,f_i),\Hpt(u_j,v_j,f_j))$.
\end{propositionA}
\begin{propositionA}\label{propA2}
The sequence of processes $\hat{\Hpt}_T^c$ is asymptotically stochastically equicontinuous, that is, for any $\varepsilon > 0$,
\[ \lim_{\delta \searrow 0} \limsup_{T \rightarrow \infty} \pr \Bigl( \sup_{|u-u^\prime|+|v-v^\prime| \atop + \dist(f,f^\prime) \le \delta} \bigl| \hat{\Hpt}_T^c(u,v,f) - \hat{\Hpt}_T^c(u^\prime,v^\prime,f^\prime) \bigr| > \varepsilon \Bigr) = 0. \]
\end{propositionA}
To prove these two results, we make use of the notation
\begin{equation}\label{HT}
\hat{\Hpt}_T^c(u,v,f) = \hat{G}_T(v,f) - \Bigl(\frac{v}{u}\Bigr) \ \hat{G}_T(u,f), 
\end{equation} 
where
\begin{equation}\label{GT}
\hat{G}_T(u,f) = \frac{1}{\sqrt{T}} \sum\limits_{t = 1}^{\lfloor uT \rfloor} \bigl( f(X_{t,T}) - \ex f(X_{t,T}) \bigr). 
\end{equation}
Combining Propositions \ref{propA1} and \ref{propA2}, the statement \eqref{theo1-1} follows from a standard functional central limit theorem (see \cite{vanderVaart1996}).
\vspace{10pt}

\textbf{Proof of Proposition \ref{propA1}.} The proof proceeds in two steps. In the first, we calculate the asymptotic covariances of the process $\hat{\Hpt}_T^c$, which is achieved by exploiting the locally stationary structure of the model variables. In the second, we apply a central limit theorem for mixing arrays. The details can be found in the Supplementary Material. \qed
\vspace{10pt}

\textbf{Proof of Proposition \ref{propA2}.} Straightforward calculations show that
\begin{align*}
\sup_{|u-u^\prime|+|v-v^\prime| \atop +\dist(f,f^\prime) \le \delta} \big| \hat{\Hpt}_T^c(u,v,f) - \hat{\Hpt}_T^c(u^\prime, v^\prime, f^\prime) \big| 
 & \le 2 \sup_{|u-u^\prime|\le \delta \atop f \in \mathcal{F}} \big| \hat{G}_T (u,f) - \hat{G}_T (u^\prime,f) \big| \\
 & \quad + 2 \sup_{\dist(f,f^\prime) \le \delta \atop u \in [0,1]} \big| \hat{G}_T (u,f) - \hat{G}_T (u,f^\prime) \big| \\
 & \quad + 2 \sup_{u \in [0,1] \atop f \in \mathcal{F}} \big| \delta^{\frac{1}{2}-\eta} \ \hat{G}_T (u,f) \big| 
   + 2 \sup_{u \in [0,\delta^{1/2 + \eta}] \atop f \in \mathcal{F}} \big| \hat{G}_T (u,f) \big|
\end{align*}
for some small $\eta > 0$. Therefore, stochastic equicontinuity follows from the statements
\begin{align}
 & \lim_{\delta\searrow 0} \limsup_{T \to \infty} \pr \Bigl( \sup_{|u-u^\prime| \leq \delta \atop f \in \mathcal{F}} \Bigl| \hat{G}_T (u,f) - \hat{G}_T (u^\prime,f) \Bigr| > \varepsilon \Bigr) = 0 \label{ec1} \\
 & \lim_{\delta\searrow 0} \limsup_{T \to \infty} \pr \Bigl( \sup_{\dist(f,f^\prime) \leq \delta \atop u \in [0,1]} \Bigl| \hat{G}_T(u,f) - \hat{G}_T (u,f^\prime) \Bigr| > \varepsilon \Bigr) = 0 \label{ec2} \\
 & \lim_{\delta\searrow 0} \limsup_{T \to \infty} \pr \Bigl( \sup_{u \in [0,1] \atop f \in \mathcal{F}} \big| \delta^{\frac{1}{2}-\eta} \ \hat{G}_T (u,f) \big| > \varepsilon  \Bigr) = 0 \label{ec3} \\
 & \lim_{\delta\searrow 0} \limsup_{T \to \infty} \pr \Bigl( \sup_{u \in [0,\delta^{1/2 + \eta}] \atop f \in \mathcal{F}} \big| \hat{G}_T (u,f) \big| > \varepsilon  \Bigr) = 0. \label{ec4}
\end{align}
\eqref{ec1}--\eqref{ec4} can be shown by very similar arguments. We thus restrict ourselves to the proof of \eqref{ec1}.

First of all, observe that for any function $g: [0,1] \to \reals$, the inequality
\begin{align*}
\sup_{|u-u^\prime| \leq \delta \atop u,u^\prime \in [0,1]} | g(u) - g(u^\prime) | 
 & \leq \max_{j=1,\ldots,\lceil 1/ \delta \rceil} \sup_{u \in [u_{j-1},u_j]} | g(u) - g(u_j) | \\
 & \quad + \max_{j=1,\ldots,\lceil 1/ \delta \rceil} \sup_{u^\prime \in [u_{j-2},u_{j+1}]} | g(u^\prime) - g(u_j) | 
\end{align*}
holds, where $u_{-1}=u_0=0$, $u_j = j\delta \ (j=1,\dots, \lceil 1/\delta \rceil - 1)$ and $u_{\lceil 1 / \delta \rceil} = u_{\lceil 1/ \delta \rceil+1}=1$. From this, it is easily seen that \eqref{ec1} is a consequence of
\begin{equation} \label{ec1A}
\lim_{\delta \searrow 0} \limsup_{T \to \infty} \pr \Bigl( \max_{j=1,\dots,\lceil 1/ \delta \rceil} \sup_{u \in [u_{j-1}, u_j]} \sup_{f \in \mathcal{F}} \Bigl| \hat{G}_T (u,f) - \hat{G}_T (j \delta,f) \Bigr| > \varepsilon \Bigr) = 0.
\end{equation}
In the sequel, we derive a suitable bound for the probability
\[ P_T(\delta,\varepsilon) =  \pr \Bigl( \max_{j=1,\dots,\lceil 1/ \delta \rceil} \sup_{u \in [u_{j-1}, u_j]} \sup_{f \in \mathcal{F}} \Bigl| \hat{G}_T (u,f) - \hat{G}_T (j \delta,f) \Bigr| > \varepsilon \Bigr) \]
in \eqref{ec1A}. To start with, we crudely bound this probability by $P_T(\delta,\varepsilon) \le \sum\nolimits_{j=1}^{\lceil 1/\delta \rceil} P_{T,j}(\delta,\varepsilon)$, where
\begin{align*}
P_{T,j}(\delta,\varepsilon) 
 & = \pr \Bigl( \sup_{u \in [u_{j-1},u_j]} \sup_{f \in \mathcal{F}} \Bigl| \hat{G}_T(u,f) - \hat{G}_T(j\delta,f) \Bigr| > \varepsilon \Bigr) \\
 & = \pr \Bigl( \max_{\lfloor (j-1)\delta T \rfloor \leq \ell \leq \lfloor j\delta T \rfloor} \sup_{f \in \mathcal{F}} \Bigl| \hat{G}_T \Big(\frac{\ell}{T},f\Big) - \hat{G}_T(j\delta,f) \Bigr| > \varepsilon \Bigr).
\end{align*} 
To bound the probabilities $P_{T,j}(\delta,\varepsilon)$, we write
\[ \hat{G}_T(j\delta,f) - \hat{G}_T \Big(\frac{\ell}{T},f\Big) = B_T^{\ell+}(f) + \sum_{r= \lceil \frac{\ell}{q} \rceil +1}^{\lfloor \frac{j \delta T}{q} \rfloor} B_{r,T}(f) + B_T^{j-}(f). \]
Here, $B_{r,T}(f)$ are blocks of length $q$ given by 
\begin{equation*} 
B_{r,T}(f) = \frac{1}{\sqrt{T}} \sum_{t=(r-1)q+1}^{rq} \bigl( f(X_{t,T}) - \ex f(X_{t,T}) \bigr), 
\end{equation*}
where as in the subsection on auxiliary results, we set $q = CT^b$ for some small $b > 0$ (specifically, $b < \frac{1}{4}$). In addition,
\begin{align*}
B_T^{\ell+}(f) &  =\frac{1}{\sqrt{T}} \sum_{t=\ell+1}^{\lceil \frac{\ell}{q} \rceil q} \bigl(f(X_{t,T}) - \ex f(X_{t,T}) \big) \\
B_T^{j-}(f)   & = \frac{1}{\sqrt{T}} \sum_{t= \lfloor \frac{j \delta T}{q} \rfloor q + 1}^{\lfloor j \delta T \rfloor} \big(f(X_{t,T}) - \ex f(X_{t,T}) \big) 
\end{align*}
denote the first and the last block, respectively. With this notation at hand, we obtain 
\begin{align*}
P_{T,j}(\delta,6\varepsilon) 
 & \leq \pr \Bigl( \max_{\lfloor (j-1)\delta T \rfloor \leq \ell \leq \lfloor j\delta T \rfloor} \sup_{f \in \mathcal{F}} \Bigl| \sum_{r=\lceil \frac{\ell}{q} \rceil + 1}^{\lfloor \frac{j \delta T}{q} \rfloor} B_{r,T}(f) \Bigr| > 4\varepsilon \Bigr) \\
 & \quad + \pr \Bigl( \max_{\lfloor (j-1)\delta T \rfloor \leq \ell \leq \lfloor j\delta T \rfloor} \sup_{f \in \mathcal{F}} |B_T^{\ell+}(f)| > \varepsilon \Bigr) + \pr \Bigl(\sup_{f \in \mathcal{F}} |B_T^{j-}(f)| > \varepsilon \Bigr)  \\
 & =: P_{T,j,1}(\delta,4\varepsilon) + P_{T,j,2}(\delta,\varepsilon) + P_{T,j,3}(\delta,\varepsilon).
\end{align*}
The terms $P_{T,j,2}$ and $P_{T,j,3}$ can be bounded by fairly straightforward arguments: Applying a maximal inequality (see e.g.\ Section 2.1.3 in \cite{vanderVaart1996}), we get that
\[ \Bigl\| \max_{\lfloor (j-1)\delta T \rfloor \leq \ell \leq \lfloor j\delta T \rfloor} \sup_{f \in \mathcal{F}} |B_T^{\ell+}(f)| \Bigr\|_p \le C (\delta T)^{1/p}  \max_{\lfloor (j-1)\delta T \rfloor \leq \ell \leq \lfloor j\delta T \rfloor} \bigl\| \sup_{f \in \mathcal{F}} |B_T^{\ell+}(f)| \bigr\|_p. \]
Moreover,
\[ \sup_{f \in \mathcal{F}} |B_T^{\ell+}(f)| \le \frac{2}{\sqrt{T}} \sum_{t=\ell+1}^{\lceil \frac{\ell}{q} \rceil q} F(X_{t,T}) \]
and by the moment conditions on the envelope $F$ in (C\ref{A3}), $\| \sup_{f \in \mathcal{F}} |B_T^{\ell+}(f)| \|_p \le C q/\sqrt{T}$. Hence by Markov's inequality, 
\[ P_{T,j,2}(\delta,\varepsilon) \le \varepsilon^{-p} \Bigl\| \max_{\lfloor (j-1)\delta T \rfloor \leq \ell \leq \lfloor j\delta T \rfloor} \sup_{f \in \mathcal{F}} |B_T^{\ell+}(f)| \Bigr\|_p^p \le C \delta T \Bigl(\frac{q}{\varepsilon \sqrt{T}}\Bigr)^p = o(1) \]
for $T \rightarrow \infty$ given that $q = T^b$ with $b < \frac{1}{4}$. By similar considerations, $P_{T,j,3}(\delta,\varepsilon)$ is seen to converge to zero as well. To deal with $P_{T,j,1}$, we split it up into two parts:
\[ P_{T,j,1}(\delta,4\varepsilon) \le \Delta_T^{(0)} + \Delta_T^{(1)} \]
with 
\begin{align*}
\Delta_T^{(0)} & = \pr \Bigl( \max_{\lfloor \frac {(j-1)\delta T}{2q} \rfloor \leq k \leq \lceil \frac {j \delta T}{2q} \rceil} \sup_{f \in \mathcal{F}} \Bigl| \sum_{r=k}^{\lfloor \frac {j \delta T}{2q} \rfloor} B_{2r,T}(f) \Bigr| > 2 \varepsilon \Bigr) \\
\Delta_T^{(1)} & = \pr \Bigl( \max_{\lfloor \frac {(j-1)\delta T}{2q} \rfloor \leq k \leq \lceil \frac {j \delta T}{2q} \rceil} \sup_{f \in \mathcal{F}} \Bigl| \sum_{r=k}^{\lceil \frac {j \delta T}{2q} \rceil} B_{2r-1,T}(f) \Bigr| > 2 \varepsilon \Bigr). 
\end{align*}
As the two terms can be treated in the same way, we restrict ourselves to $\Delta_T^{(1)}$. Applying a version of Ottaviani's inequality for $\alpha$-mixing processes (which has the form stated in Chapter 10.2 of \cite{Lin2010} and can be proven by the arguments therein), we obtain that
\begin{equation} \label{delest}
\Delta_T^{(1)} \leq \frac{\pr \Bigl( \sup\limits_{f \in \mathcal{F}} \Bigl| \sum\limits_{r= \lfloor \frac{(j-1)\delta T}{2q} \rfloor}^{\lceil \frac {j \delta T}{2q} \rceil} B_{2r-1,T}(f) \Bigr| > \varepsilon \Bigr) + \frac{\delta T}{2q} \alpha (q)}{1 - \max\limits_{\lfloor \frac {(j-1)\delta T}{2q} \rfloor \leq k \leq \lceil \frac {j \delta T}{2q} \rceil} \pr \Bigl( \sup\limits_{f \in \mathcal{F}} \Bigl| \sum\limits_{r= \lfloor \frac{(j-1)\delta T}{2q} \rfloor}^k B_{2r-1,T}(f) \Bigr| > \varepsilon \Bigr)}.
\end{equation}
In order to bound the right-hand side of \eqref{delest}, we make use of the random variables
\begin{equation*} 
Q_{r,T}(f) = \frac {1}{\sqrt{(k^\prime - k + 1)q}} \sum^{(2r-1)q \wedge T}_{t=(2r-2)q+1} \bigl( f(X_{t,T}) - \ex f(X_{t,T}) \bigr)
\end{equation*}
and $W_T(k,k^\prime) = \sup_{f \in \mathcal{F}} | \sum\nolimits_{r=k}^{k^\prime} Q_{r,T}(f) |$, which have been introduced at the beginning of the appendix. Combining Lemmas \ref{lemmaA1} and \ref{lemmaA2} and noting that $\int_0^{\textnormal{diam}(\mathcal{F})} \mathcal{N}(w/2,\mathcal{F},d)^{1/p} dw$ is finite by assumption (C\ref{A3}), we get that $\ex \bigl[ | W_T (k,k^\prime) |^p \bigr] \leq C < \infty$ for some sufficiently large constant $C$. This implies that
\begin{align*}
\pr \Bigl( \sup_{f \in \mathcal{F}} \Bigl| \sum_{r=k}^{k^\prime} B_{2r-1,T}(f) \Bigr| > \varepsilon \Bigr) 
 & = \pr \Bigl( W_T(k,k^\prime) > \frac{\varepsilon \sqrt{T}}{\sqrt{(k^\prime - k + 1)q}} \Bigr) \nonumber \\
 & \leq \ex \bigl[ | W_T(k,k^\prime)|^p \bigr] \Bigl( \frac{(k^\prime - k + 1)q}{\varepsilon^2T} \Bigr)^{p/2} 
   \leq C \Bigl( \frac{(k^\prime - k + 1)q}{\varepsilon^2T} \Bigr)^{p/2}. 
\end{align*}
Specifically, whenever $(k - k^\prime + 1)q \leq \delta T$, 
\begin{equation}\label{delest1}
\pr \Bigl( \sup_{f \in \mathcal{F}} \Bigl| \sum_{r=k}^{k^\prime} B_{2r-1,T}(f) \Bigr| > \varepsilon \Bigr) \leq C \frac {\delta^{p/2}}{\varepsilon^p}. 
\end{equation}
With (\ref{delest1}), it is easy to see that the denominator in \eqref{delest} is bounded away from zero as $T \to \infty$ and to infer that
\[ \Delta_T^{(1)} \leq C \Bigl( \frac{\delta^{p/2}}{\varepsilon^p} + \frac{\delta T}{2q} \alpha (q) \Bigr). \]
Using an analogous bound for the term $\Delta_T^{(0)}$, it follows that
\[  P_T(\delta,\varepsilon) \le \sum_{j=1}^{\lceil 1/ \delta \rceil} P_{T,j}(\delta,\varepsilon) \leq C \Bigl\lceil \frac{1}{\delta} \Bigr\rceil \Bigl( \frac{\delta^{p/2}}{\varepsilon^p} + \frac{\delta T}{2q} \alpha (q) \Bigr). \]
This yields that $\lim_{\delta \searrow 0} \limsup_{T \to \infty} P_T(\delta,\varepsilon) = 0$ and the assertion \eqref{ec1A} follows. By the discussion at the beginning of this proof we obtain \eqref{ec1}, which implies stochastic equicontinuity. \qed

\subsection*{Proof of Theorem \ref{theo-convergence}}

The proof is an immediate consequence of the following two statements:
\begin{align}
 & \pr \bigl(\hat{u}_0(\tau_T) < u_0 \bigr) = o(1) \label{conv-proof-1} \\
 & \pr \bigl( \hat{u}_0(\tau_T) > u_0 + K \gamma_T \bigr) = o(1) \label{conv-proof-2}
\end{align} 
for some sufficiently large constant $K > 0$. 
\vspace{10pt}

\textbf{Proof of (\ref{conv-proof-1}).} It holds that 
\begin{align*}
\pr \bigl( \hat{u}_0(\tau_T) < u_0 \bigr)
 & \le \pr \Bigl( \sqrt{T} \hat{\Dsup}_T(u) > \tau_T \text{ for some } u < u_0 \Bigr) \\
 & \le \pr \Bigl( \sqrt{T} \Dsup(u) + \hat{\Hsup}_T(u) > \tau_T \text{ for some } u < u_0 \Bigr) 
   \le \pr \Bigl( \sup_{u \in [0,1]} \hat{\Hsup}_T(u) > \tau_T \Bigr),
\end{align*}
where the second inequality follows from the fact that $\sqrt{T} \hat{\Dsup}_T(u) \le \sqrt{T} \Dsup(u) + \hat{\Hsup}_T(u)$ and the third one exploits the fact that $\Dsup(u) = 0$ at points $u < u_0$. From Corollary \ref{corollary-measure}, we know that $\sup_{u \in [0,1]} \hat{\Hsup}_T(u) = \hat{\Hmax}_T(1)$ converges in distribution to $\Hmax(1)$. Moreover, the distribution function $F$ of $\Hmax(1)$ is continuous on $[0,\infty)$ by the results of Section 3 in \cite{Lifshits1982}. We can thus infer that the distribution function $F_T$ of $\hat{\Hmax}_T(1)$ uniformly converges to $F$ on $[0,\infty)$. As a result, we obtain that
\begin{align*} 
\pr \bigl( \hat{\Hmax}_T(1) > \tau_T \bigr)
 & = 1 - F_T(\tau_T) = [1 - F(\tau_T)] + [F(\tau_T) - F_T(\tau_T)] = o(1),
\end{align*}
which in turn yields \eqref{conv-proof-1}. \qed 
\vspace{10pt}

\textbf{Proof of (\ref{conv-proof-2}).} Similarly as above, we can write
\begin{align*}
\pr \bigl( \hat{u}_0(\tau_T) > u_0 + K \gamma_T \bigr)
 & \le \pr \Bigl( \sqrt{T} \hat{\Dsup}_T(u) \le \tau_T \text{ for some } u > u_0 + K \gamma_T \Bigr) \\
 & \le \pr \Bigl( \sqrt{T} \Dsup(u) - \hat{\Hsup}_T(u) \le \tau_T \text{ for some } u > u_0 + K \gamma_T \Bigr), 
\end{align*}
the last line following from the fact that $\sqrt{T} \Dsup(u) - \hat{\Hsup}_T(u) \le \sqrt{T} \hat{\Dsup}_T(u)$. Next notice that 
\[ \min_{u \in [u_0 + K \gamma_T,1]} \Dsup(u) \ge \frac{c_{\kappa} (K\gamma_T)^{\kappa}}{2} \]
for sufficiently large $T$, which easily follows upon inspection of \eqref{smoothness-D}. This allows us to infer that
\begin{align*}
 & \pr \Bigl( \sqrt{T} \Dsup(u) - \hat{\Hsup}_T(u) \le \tau_T \text{ for some } u > u_0 + K \gamma_T \Bigr) \\
 & \le \pr \Bigl( \frac{\sqrt{T} c_{\kappa} (K\gamma_T)^{\kappa}}{2} - \hat{\Hmax}_T(1) \le \tau_T \Bigr) \\
 & \le \pr \Bigl( \frac{\sqrt{T} c_{\kappa} (K\gamma_T)^{\kappa}}{2} - \hat{\Hmax}_T(1) \le \tau_T, \hat{\Hmax}_T(1) \le b_T \Bigr) + \pr \Bigl( \hat{\Hmax}_T(1) > b_T \Bigr) 
   =: P_1 + P_2,
\end{align*}
where $b_T$ is some diverging sequence of positive numbers satisfying $b_T / \tau_T \rightarrow 0$. As already seen in the proof of \eqref{conv-proof-1}, it holds that $P_2 = o(1)$. Moreover, $P_1 = 0$ for sufficiently large $T$ if we set $\gamma_T = (\tau_T/\sqrt{T})^{1/\kappa}$ and choose $K$ to be sufficiently large. This shows \eqref{conv-proof-2}. \qed

\subsection*{Proof of Theorem \ref{theo-threshold} and Corollary \ref{corollary-threshold}}

Due to space constraints, the proofs are deferred to the Supplementary Material.

\setlength{\bibsep}{1pt}
\begin{small}
\bibliographystyle{ims}
\bibliography{bibliography}
\end{small}

\end{document}


\headin{Supplementary Material for}{``Detecting Gradual Changes in Locally}{Stationary Processes''}
\renewcommand{\thefootnote}{*}
\authors{Michael Vogt}{University of Konstanz}{Holger Dette}{Ruhr-Universit\"at Bochum}
\version{\today}

\renewcommand{\thefootnote}{\arabic{footnote}}
\setcounter{footnote}{0}

\begin{abstract}
\noindent
In this supplement, we examine the finite sample performance of our method by further simulations. In addition, we provide the technical details and proofs that are omitted in the paper.
\end{abstract}

\renewcommand{\baselinestretch}{1.15}\normalsize
\def\theequation{S.\arabic{equation}}
\setcounter{equation}{0}

\section{Simulations}

In what follows, we continue the simulation study from Section 7.1  of the paper. As announced there, we examine a volatility model together with a multivariate extension of it. The univariate model is
\begin{equation} \label{mod2}
X_{t,T} = \sigma \Bigl(\frac{t}{T}\Bigr) \varepsilon_t, 
\end{equation}
where $\sigma$ is a time-varying volatility function and $\varepsilon_t$ are i.i.d.\ residuals that are normally distributed with zero mean and unit variance. This is the same model as discussed in the application on the S\&P 500 returns in Section 7.3 of the paper. Our aim is to estimate the time point  where the volatility function $\sigma$ starts to vary over time. We consider two different specifications of $\sigma$,  
\begin{align*}
\sigma_1(u) & = 1(u < 0.5) + 2 \cdot 1(u \ge 0.5) \\
\sigma_2(u) & = 1(u < 0.5) + \{1 + 10(u-0.5)\} \cdot 1( 0.5 < u < 0.6) + 2 \cdot 1 (u \ge 0.6), 
\end{align*}
both of which are equal to $1$ on the interval $[0,0.5]$ and then start to vary over time. Thus, $u_0 = 0.5$ in both cases. Analogously to the time-varying mean setting, $\sigma_1$ has a jump at $u_0 = 0.5$, whereas $\sigma_2$ smoothly deviates from its baseline value $1$.

The multivariate extension of model \eqref{mod2} is given by the equation
\begin{equation} \label{mod3}
X_{t,T} = \Sigma \Bigl(\frac{t}{T}\Bigr) \varepsilon_t, 
\end{equation}
where $X_{t,T} = (X_{t,T,1},X_{t,T,2})^{^\intercal}$ are bivariate random variables, $\Sigma(u)$ is a $2 \times 2$-matrix for each time point $u$ and $\varepsilon_t = (\varepsilon_{t,1},\varepsilon_{t,2})^{^\intercal}$ are bivariate standard normal i.i.d.\ residuals. Since $\Sigma^2(\frac{t}{T}) := \Sigma(\frac{t}{T}) \Sigma^{^\intercal}(\frac{t}{T}) = \ex[X_{t,T} X_{t,T}^{^\intercal}]$, the time-varying matrix $\Sigma^2(\frac{t}{T})$ is the covariance matrix of $X_{t,T}$. Our aim is to estimate the time point where this matrix starts to vary over time. Put differently, we want to localize the time point where the covariance structure of $X_{t,T}$ starts to change. The stochastic feature of interest is thus the vector of covariances $\lambda_{t,T} = (\nu_{t,T}^{(1,1)}, \nu_{t,T}^{(1,2)}, \nu_{t,T}^{(2,2)})^{^\intercal}$, where $\nu_{t,T}^{(i,j)} = \ex[X_{t,T,i} X_{t,T,j}]$. We consider two different specifications of the volatility matrix $\Sigma$, 
\begin{align*}
\Sigma_1(u) & = \sigma_1(u) \cdot A \\
\Sigma_2(u) & = \sigma_2(u) \cdot A, 
\end{align*}
where 
\[ A A^{^\intercal} = \left( \begin{matrix} 1 & 0.5 \\ 0.5 & 1 \end{matrix} \right), \quad \text{or put differently,} \quad A \approx \left( \begin{matrix} 0.87 & -0.5 \\ 0.87 & \phantom{ - } 0.5 \end{matrix} \right) \]
and $\sigma_1(u)$ along with $\sigma_2(u)$ are defined above. Both matrices $\Sigma_1(u)$ and $\Sigma_2(u)$ are constant on the interval $[0,0.5]$ and then start to vary over time. Hence, as in the univariate case, $u_0 = 0.5$.

\begin{figure}[H]
\centering
\includegraphics[width=0.32\textwidth]{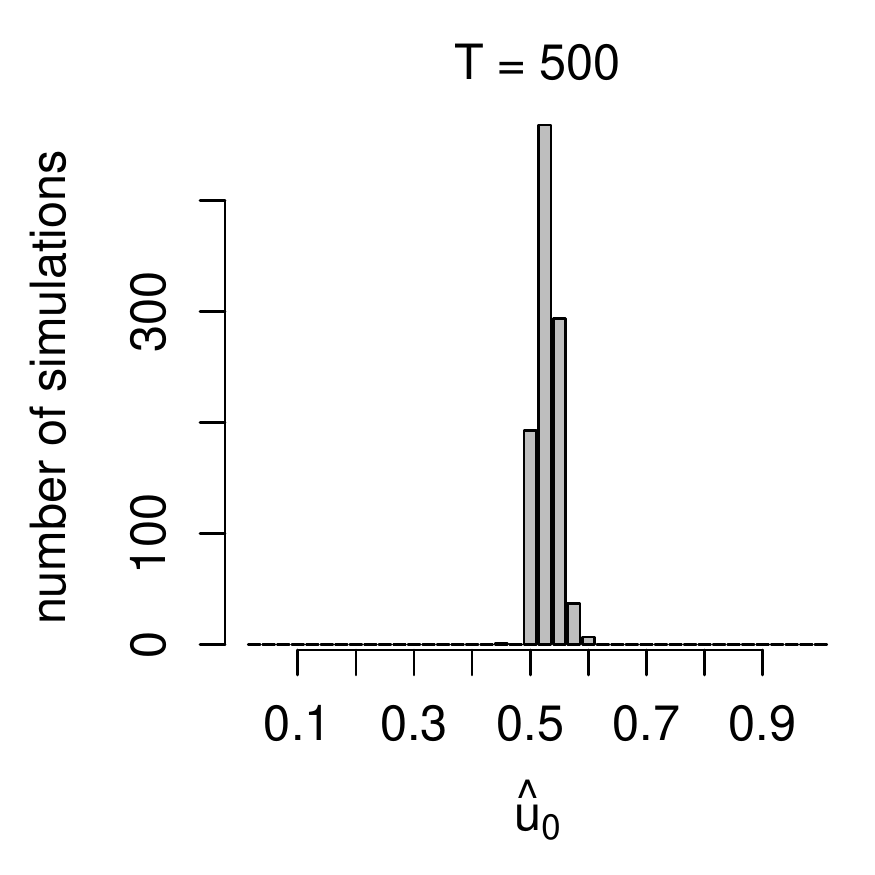}
\includegraphics[width=0.32\textwidth]{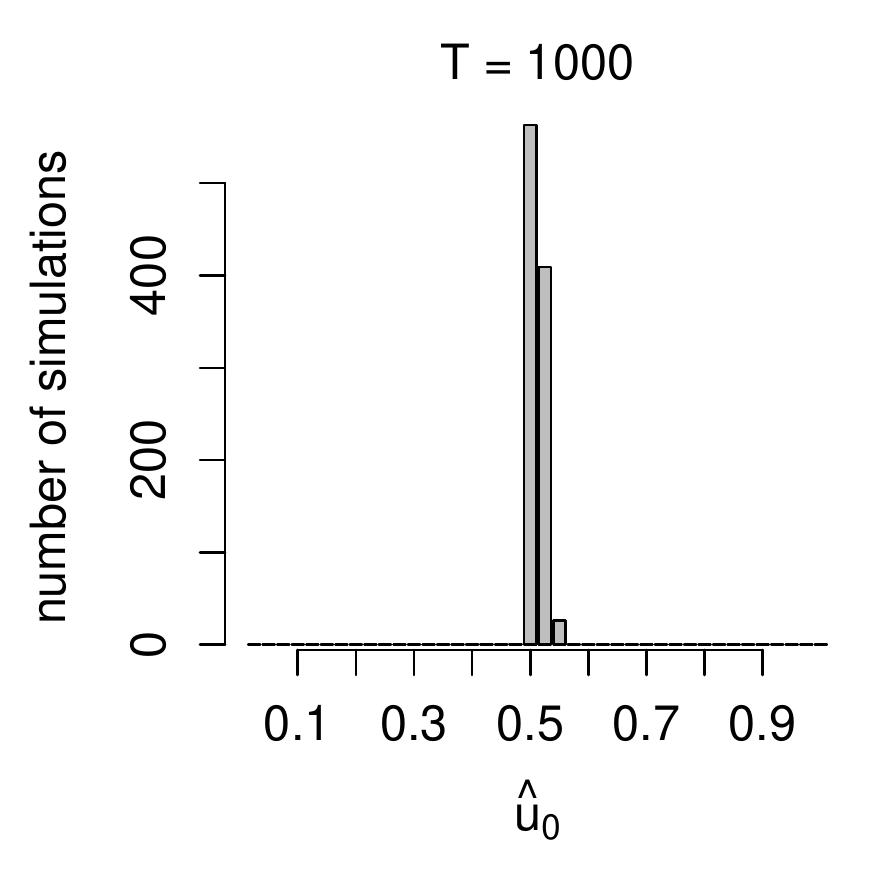} \\
\includegraphics[width=0.32\textwidth]{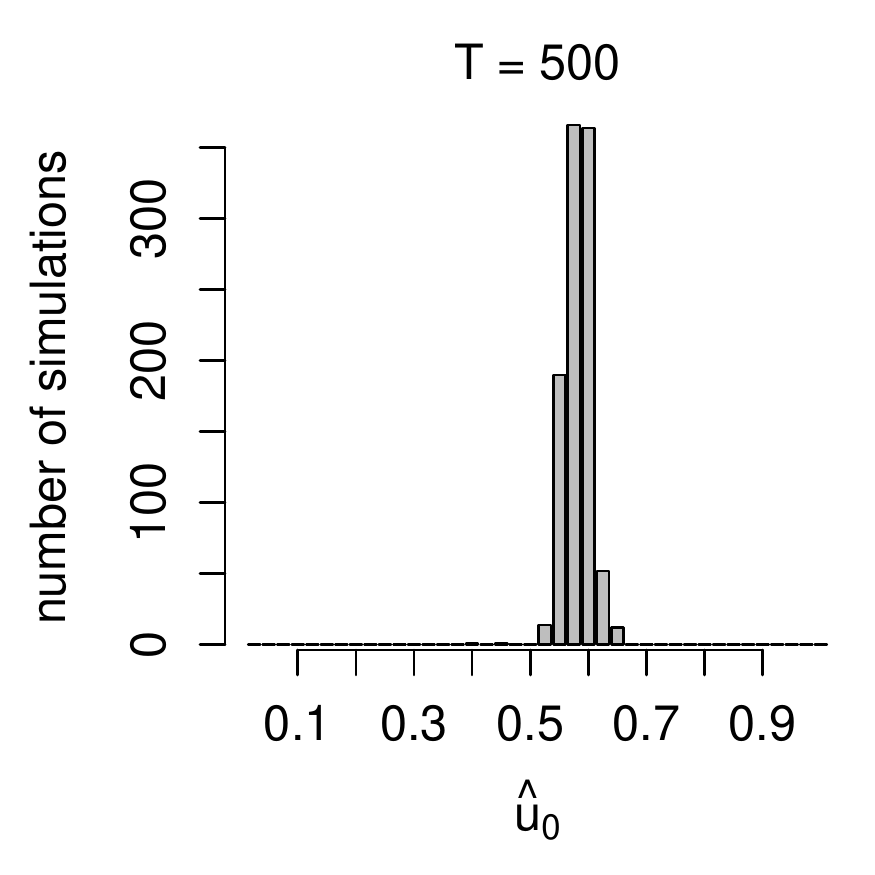}
\includegraphics[width=0.32\textwidth]{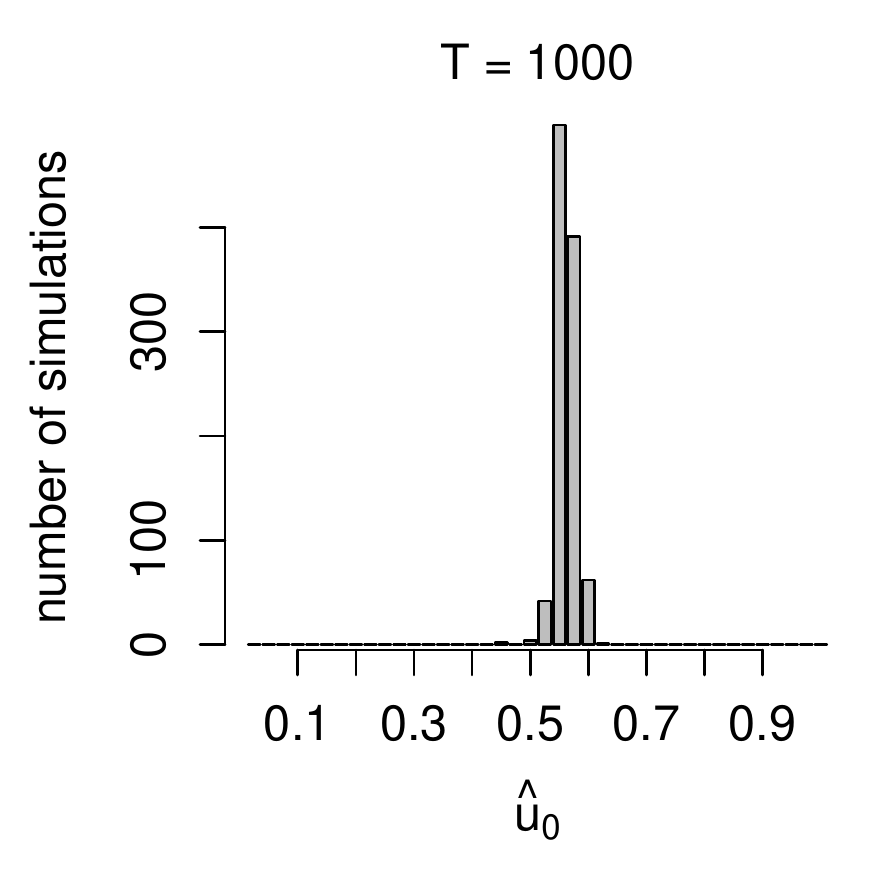}
\caption{Simulation results for model \eqref{mod2} with the volatility function $\sigma_1$ (upper panel) and the function $\sigma_2$ (lower panel).}\label{fig-sig}
\end{figure}

We implement our method as described in Section 6 of the paper, setting the parameter $\alpha$ to equal $0.1$. To calculate the quantiles in the first step of the implementation, we employ an estimator of the form (6.1) with $h=0.2$ and a bandwidth $b$ of zero, exploiting the fact that the simulated data are independent. The resulting estimator is denoted by $\hat{u}_0$. For each model specification, we draw $N=1000$ samples of length $T \in \{ 500,1000 \}$ and compute the estimate of $u_0$ for each draw. The results are presented by means of histograms in the same way as in Section 7 of the paper.

\begin{figure}[H]
\centering
\includegraphics[width=0.6\textwidth]{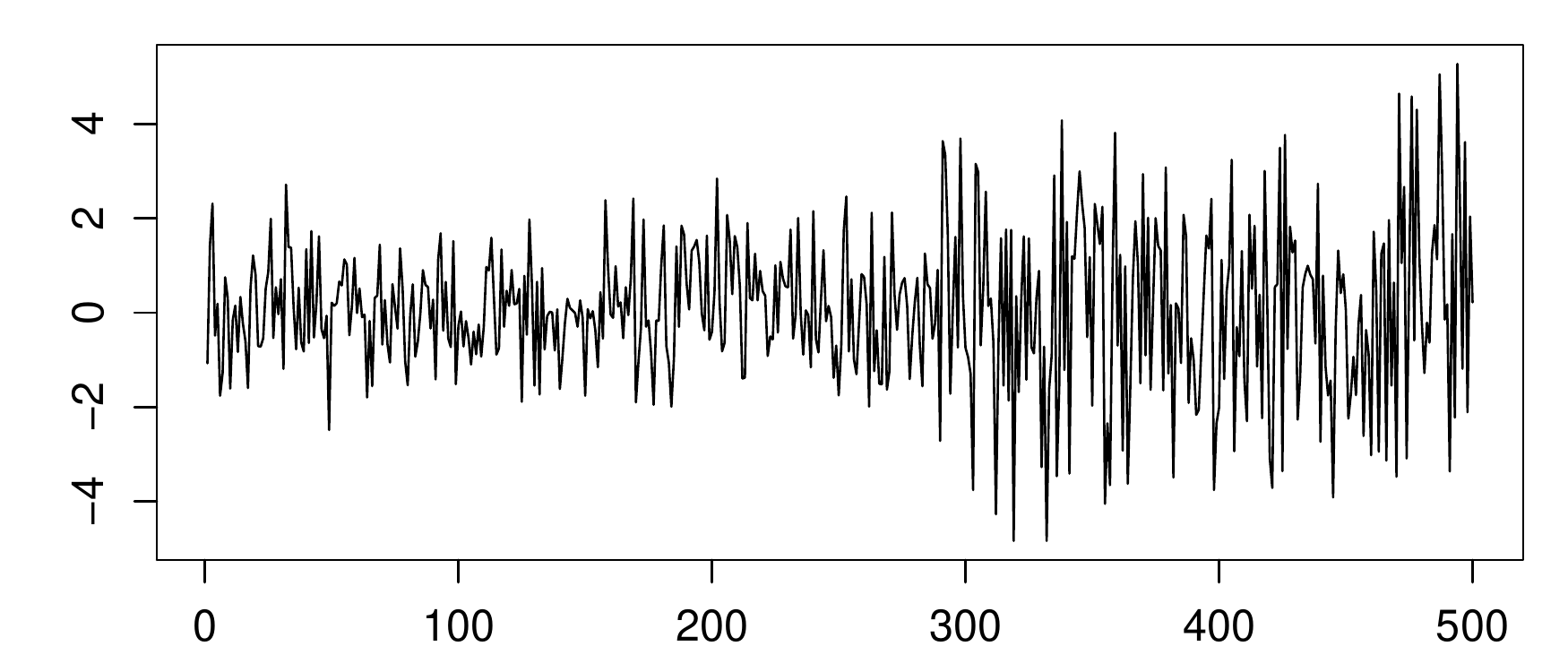}
\caption{A typical sample path of length $T=500$ for model \eqref{mod2} with $\sigma_2$.}\label{fig-sig2-data}
\end{figure}

We first discuss the results on the univariate model \eqref{mod2}. The upper panel of Figure \ref{fig-sig} presents the histograms for the design with $\sigma_1$, the lower panel those for the design with $\sigma_2$. The results are fairly similar to those from the time-varying mean setting: Our method is again able to detect the point $u_0$ quite precisely in the jump design with $\sigma_1$. The histograms in the setup with $\sigma_2$ are a bit more dispersed, reflecting the fact that it is harder to localize a gradual change than a jump.

\begin{figure}[H]
\centering
\includegraphics[width=0.32\textwidth]{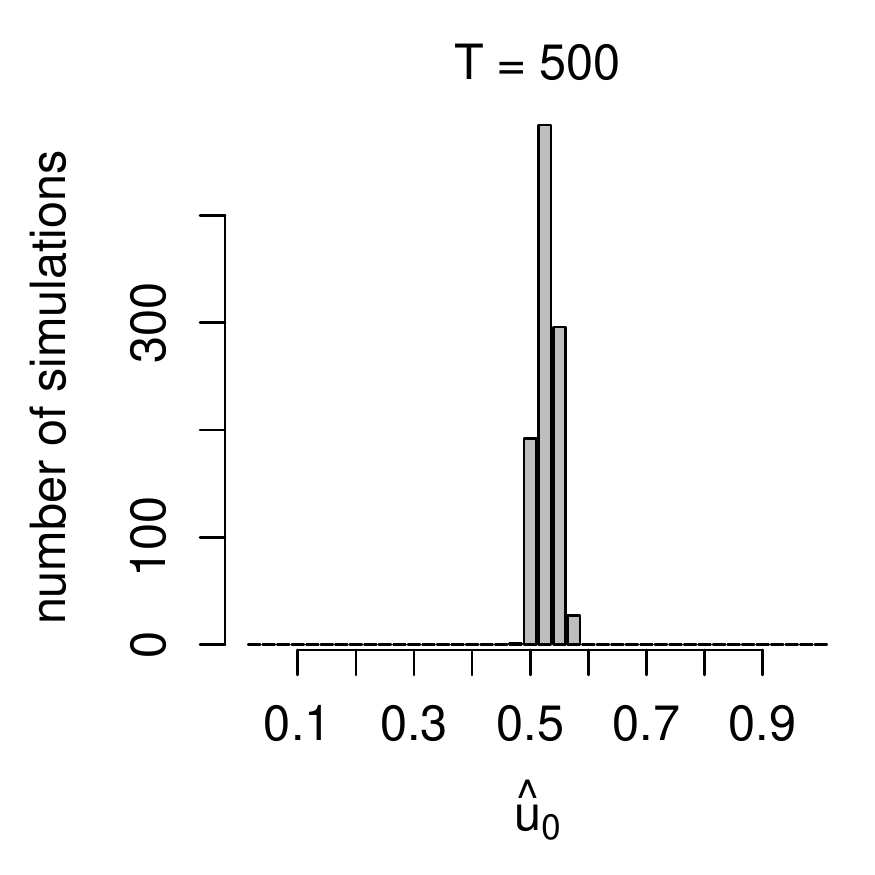}
\includegraphics[width=0.32\textwidth]{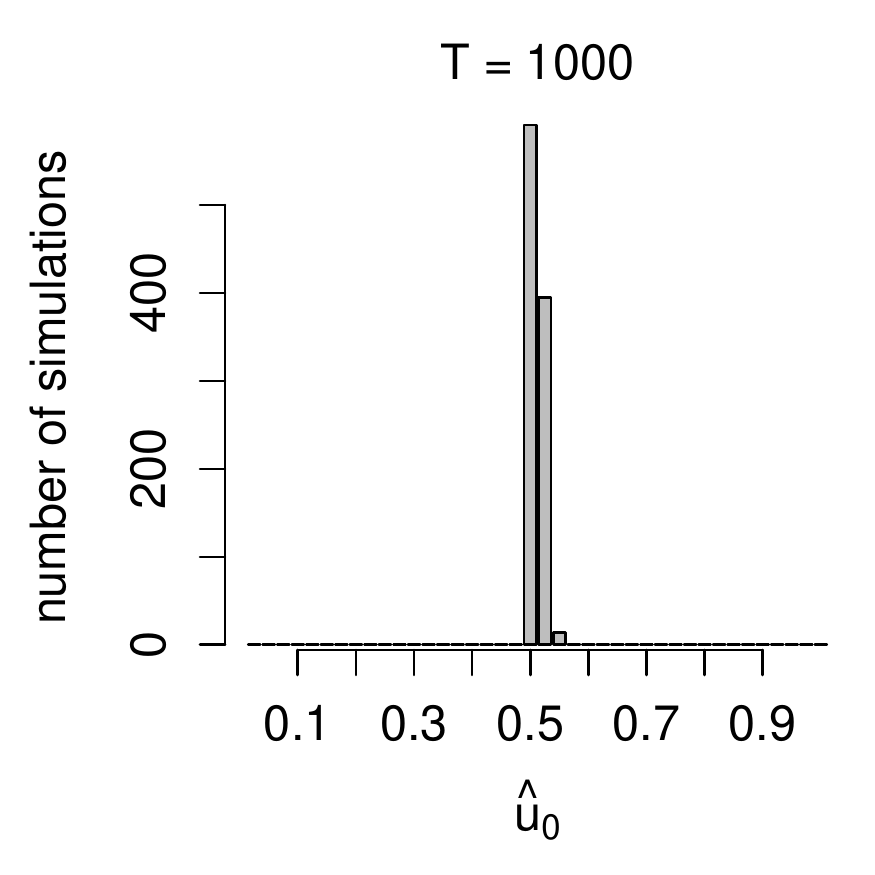} \\
\includegraphics[width=0.32\textwidth]{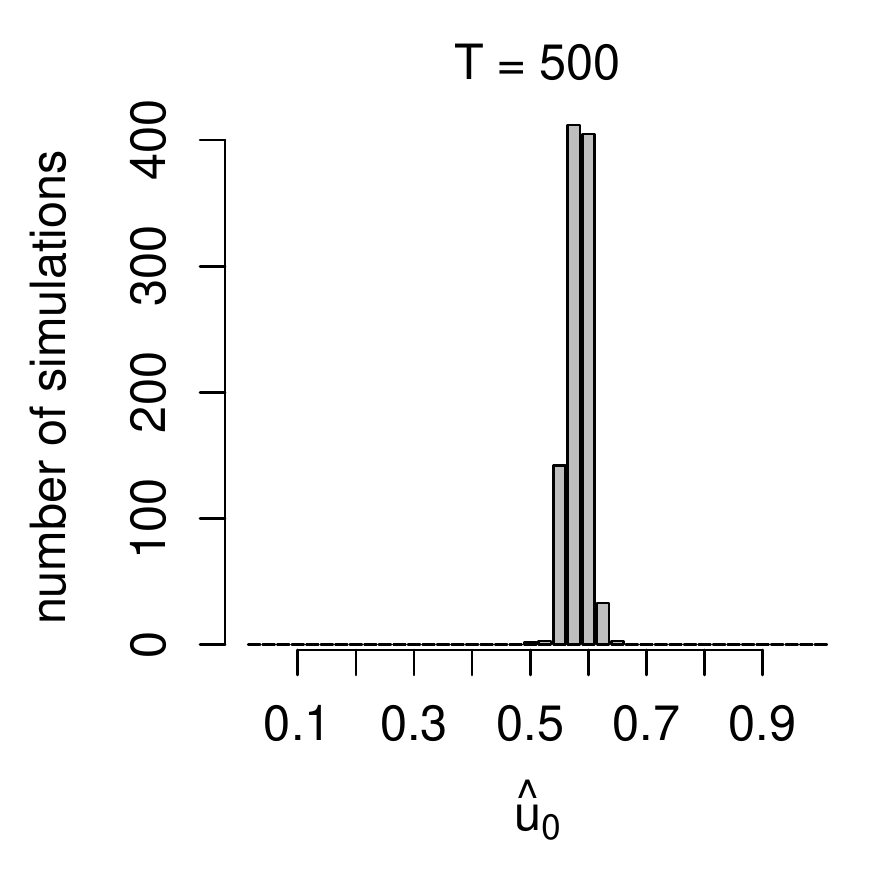}
\includegraphics[width=0.32\textwidth]{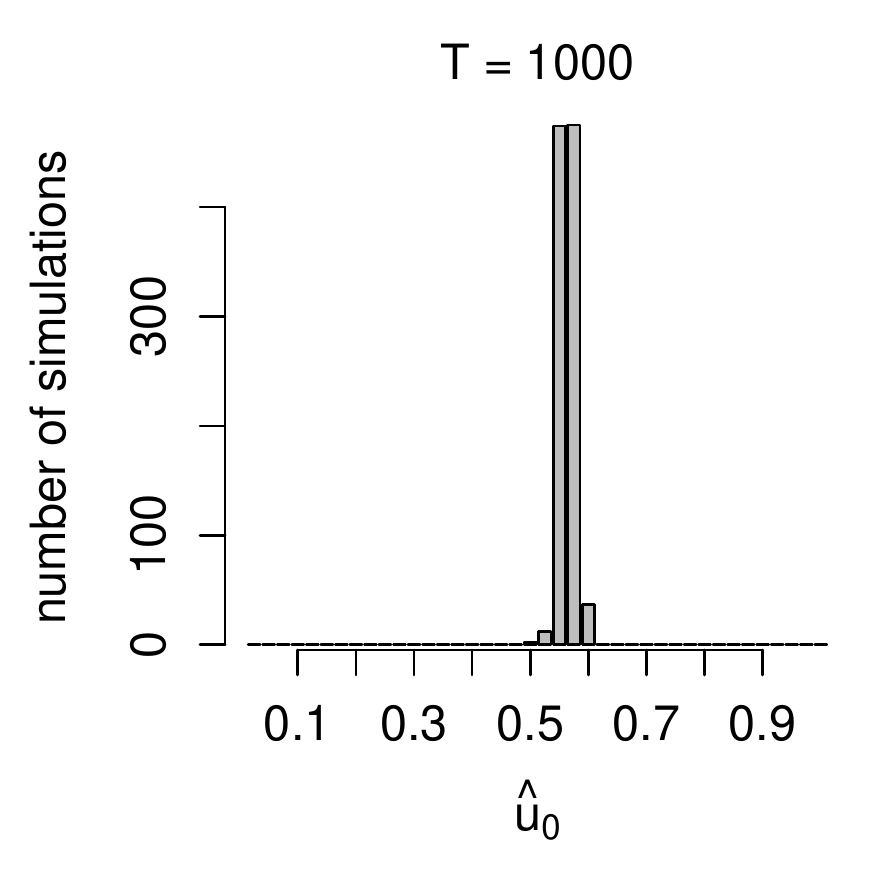}
\caption{Simulation results for model \eqref{mod3} with the volatility matrix $\Sigma_1$ (upper panel) and the matrix $\Sigma_2$ (lower panel).}\label{fig-Sig}
\end{figure}

Figure \ref{fig-sig2-data} shows a typical sample path of length $T=500$ for the design with $\sigma_2$. As can be seen, the increase in the volatility level is hardly visible close to $u_0 = 0.5$ and only becomes apparent with some delay. It is thus natural that our procedure detects the time-variation in the volatility level only with a bit of delay. This produces the upward bias in the histograms which becomes less pronounced in larger samples.

We finally turn to the results for the bivariate model \eqref{mod3}. The histograms for the model with $\Sigma_1$ are displayed in the upper panel of Figure \ref{fig-Sig}, those for the design with $\Sigma_2$ in the lower panel. Overall, the estimates give a good approximation to the true value $u_0$, those in the jump design with $\Sigma_1$ being a bit more precise than those in the  gradual  change design. Moreover, the histograms again make visible an upward bias which is comparable in size to that in the univariate setting.

\section{Technical Details}

\textbf{Proof of Proposition A.1.} We first calculate the asymptotic expectation and covariances of the process $\hat{\Hpt}_T^c$. As the process is centered, it holds that $\ex [\hat{\Hpt}_T^c(u,v,f)] = 0$. Moreover, 
\begin{align}
\cov \bigl( \hat{\Hpt}_T^c(u_1,v_1,f_1),\hat{\Hpt}_T^c(u_2,v_2,f_2) \bigr)
 & = \frac{v_1 v_2}{u_1 u_2} \ \ex \bigl[ \hat{G}_T(u_1,f_1) \hat{G}_T(u_2,f_2) \bigr] \nonumber \\
 & \quad - \frac{v_2}{u_2} \ \ex \bigl[ \hat{G}_T(v_1,f_1) \hat{G}_T(u_2,f_2) \bigr] \nonumber \\
 & \quad - \frac{v_1}{u_1} \ \ex \bigl[ \hat{G}_T(u_1,f_1) \hat{G}_T(v_2,f_2) \bigr] \nonumber \\
 & \quad + \ex \bigl[ \hat{G}_T(v_1,f_1) \hat{G}_T(v_2,f_2) \bigr]. \label{cov-1}
\end{align}
In what follows, we show that
\begin{equation}\label{cov-2}
\ex \bigl[ \hat{G}_T(u_1,f_1) \hat{G}_T(u_2,f_2) \bigr] = \sum\limits_{\ell = -\infty}^{\infty} \int_0^{\min \lbrace u_1,u_2 \rbrace} c_{\ell}(w) dw + o(1) 
\end{equation}
with $c_{\ell}(w) = c_{\ell}(w,f_1,f_2) = \cov(f_1(X_0(w)),f_2(X_{\ell}(w)))$. Plugging \eqref{cov-2} into \eqref{cov-1} yields
\[ \cov \bigl( \hat{\Hpt}_T^c(u_1,v_1,f_1),\hat{\Hpt}_T^c(u_2,v_2,f_2) \bigr) = \cov \bigl( \Hpt(u_1,v_1,f_1),\Hpt(u_2,v_2,f_2) \bigr) + o(1). \]
Hence, the covariances of $\hat{\Hpt}_T^c$ converge to those of the Gaussian process $\Hpt$.

To show \eqref{cov-2}, we assume without loss of generality that $u_1 \le u_2$. Exploiting the mixing condition (C2) by means of Davydov's inequality, it can be seen that $\cov \bigl(f_1(X_{t,T}),f_2(X_{s,T})\bigr) \le C \alpha(|s-t|) \le C a^{|s-t|}$ for some $a < 1$ and a sufficiently large constant $C$. We thus obtain that
\begin{align*}
 & \ex \bigl[ \hat{G}_T(u_1,f_1) \hat{G}_T(u_2,f_2) \bigr] \\
 & = \frac{1}{T} \sum_{t=1}^{\lfloor u_1T \rfloor} \sum_{s=1}^{\lfloor u_2T \rfloor} \cov \bigl(f_1(X_{t,T}),f_2(X_{s,T})\bigr) \\ 
 & = \frac{1}{T} \sum_{t=1}^{\lfloor u_1T \rfloor} \sum_{s=1}^{\lfloor u_2T \rfloor} I\{ |s-t| \leq C^* \log T \} \cov \bigl(f_1(X_{t,T}),f_2(X_{s,T})\bigr) + o(1) \\
 & =: Q_T^{(1)} +  Q_T^{(2)} + Q_T^{(3)} + o(1)  
\end{align*} 
for some sufficiently large constant $C^*$, where the random variables $Q_T^{(j)} \ (j=1,2,3)$ are defined by
\begin{align*}
 Q_T^{(1)} & = \frac{1}{T} \sum_{\ell = 1}^{\lceil C^* \log T \rceil} \sum_{t=1}^{T-\ell} I \bigl\{ t \le \lfloor u_1T \rfloor, t+ \ell \le \lfloor u_2 T \rfloor \bigr\} \cov \bigl( f_1(X_{t,T}),f_2(X_{t+\ell,T}) \bigr) \\
 Q_T^{(2)} & = \frac{1}{T} \sum_{t=1}^{\lfloor u_1T \rfloor} \cov \bigl( f_1(X_{t,T}),f_2(X_{t,T}) \bigr) \\
 Q_T^{(3)} & = \frac{1}{T} \sum_{\ell = 1}^{\lceil C^* \log T \rceil} \sum_{t= \ell + 1}^T I \bigl\{ t \le \lfloor u_1T \rfloor, t- \ell \le \lfloor u_2 T \rfloor \bigr\} \cov \bigl(f_1(X_{t,T}),f_2(X_{t-\ell,T}) \bigr).
\end{align*}
By assumption (C4), it follows for $\ell \leq \lceil C^* \log T \rceil$ and any $w$ with $|w-\frac{t}{T}| \leq \frac{1}{T}$ that
\begin{align*}
c_{t,T,\ell} 
 & := \cov \bigl( f_1(X_{t,T}),f_2(X_{t+\ell,T}) \bigr) \\
 & \phantom{:}= \cov \Bigl( f_1\Big(X_t\Big(\frac{t}{T}\Big)\Big),f_2\Big(X_{t+\ell}\Big(\frac{t+\ell}{T}\Big)\Big) \Bigr) + O\Bigl(\frac{\log T}{T}\Bigr) \\
 & \phantom{:}= \cov \Bigl( f_1\Big(X_t\Big(\frac{t}{T}\Big)\Big),f_2\Big(X_{t+\ell}\Big(\frac{t}{T}\Big)\Big) \Bigr) + O\Bigl(\frac{\log T}{T}\Bigr) \\
 & \phantom{:}= \cov \bigl( f_1(X_0(w)),f_2(X_{\ell}(w)) \bigr) + O \Bigl(\frac{\log T}{T}\Bigr) \\
 & \phantom{:}=: c_{\ell}(w) + O \Bigl( \frac{\log T}{T} \Bigr),
\end{align*}
the last line defining $c_{\ell}(w)$ in an obvious manner. From this, it is easy to see that
\begin{align*}
\frac{1}{T} \sum_{\ell = 1}^{\lceil C^* \log T \rceil} \sum_{t=1}^{T-\ell} |c_{t,T,\ell}| 
 & = \sum_{\ell = 1}^{\lceil C^* \log T \rceil} \sum_{t=1}^{T - \ell} \int_{\frac{t-1}{T}}^{\frac{t}{T}} \Big|c_{\ell}\Bigl(\frac{t}{T}\Bigr)\Big| dw + O \Bigl( \frac{(\log T)^2}{T} \Bigr) \\
 & = \sum_{\ell = 1}^{\lceil C^* \log T \rceil} \sum_{t=1}^{T - \ell} \int_{\frac{t-1}{T}}^{\frac{t}{T}} |c_{\ell}(w)| dw + O \Bigl( \frac{(\log T)^2}{T} \Bigr) \\
 & = \sum_{\ell = 1}^{\lceil C^* \log T \rceil} \int_0^1 |c_\ell(w)| dw + O \Bigl( \frac{(\log T)^2}{T} \Bigr).
\end{align*}
Because of the mixing assumption (C2), the left-hand side of this equation is bounded as $T \to \infty$ and consequently $\sum_{\ell = 1}^\infty \int^1_0 c_{\ell}(w)dw$ is absolutely convergent. Therefore we obtain for the term $Q_T^{(1)}$ as $T \to \infty$ (recall that $u_1 \le u_2$) 
\begin{align*}
Q_T^{(1)} 
 & = \sum_{\ell = 1}^{\lceil C^* \log T \rceil} \sum_{t=1}^{\lfloor u_1 T \rfloor - \ell} \int_{\frac{t-1}{T}}^{\frac{t}{T}} c_{\ell}(w) dw + O \Bigl( \frac{(\log T)^2}{T} \Bigr) \\
 & = \sum_{\ell = 1}^\infty \int_0^{u_1} c_{\ell}(w) dw + O \Bigl( \frac{(\log T)^2}{T} \Bigr)
\end{align*}
and similarly
\[ Q_T^{(2)} = \int_0^{u_1} c_0(w) dw + O \Bigl( \frac{(\log T)^2}{T} \Bigr), \quad Q_T^{(3)} = \sum_{\ell = 1}^\infty \int_0^{u_1} c_{-\ell}(w) dw + O \Bigl( \frac{(\log T)^2}{T} \Bigr). \]
Putting everything together, we arrive at \eqref{cov-2}.

Having calculated the asymptotic covariance structure of $\hat{\Hpt}_T^c$, we now apply a central limit theorem for mixing arrays of random variables (see e.g.\ \cite{Liebscher1996}) together with the Cram\'{e}r-Wold device to obtain weak convergence of the finite dimensional distributions. \qed
\vspace{10pt}

\textbf{Proof of Theorem 5.4.} We first derive (5.7) which says that 
\[ \pr \bigl( \hat{u}_0(\tau_{\alpha}) < u_0 \bigr) \le \alpha + o(1). \]
It holds that
\begin{align*}
\pr \bigl( \hat{u}_0(\tau_{\alpha}) < u_0 \bigr)
 & \le \pr \bigl( \sqrt{T} \hat{\Dsup}_T(u) > \tau_{\alpha} \text{ for some } u < u_0 \bigr) \\
 & \le \pr \bigl( \sqrt{T} \hat{\Dmax}_T(u_0) > \tau_{\alpha} \bigr) \\
 & = \pr \bigl( \hat{\Hmax}_T(u_0) > \tau_{\alpha} \bigr). 
\end{align*}
We now make use of the following fact which is a direct consequence of the results from Section 3 in \cite{Lifshits1982}:
\begin{enumerate}[label=($\ast$),leftmargin=0.75cm]
\item \label{fact1} For each $u$, the random variable 
\[ \Hmax(u) = \sup_{f \in \mathcal{F}} \sup_{0 \le w \le v \le u} | H(v,w,f)| \]
has a distribution function which is continuous on $[0,\infty)$. 
\end{enumerate}
By \ref{fact1}, we obtain that
\begin{align*}
\pr \bigl( \hat{\Hmax}_T(u_0) > \tau_{\alpha} \bigr) 
 & = \pr \bigl( \Hmax(u_0) > \tau_{\alpha} \bigr) + \Big[ \pr \bigl( \hat{\Hmax}_T(u_0) > \tau_{\alpha} \bigr) - \pr \bigl( \Hmax(u_0) > \tau_{\alpha} \bigr) \Big] \\
 & = \pr \bigl( \Hmax(u_0) > \tau_{\alpha} \bigr) + o(1) = \alpha + o(1),
\end{align*}
where the last equality is due to the fact that $\tau_{\alpha} = q_{\alpha}(u_0)$ is the $(1-\alpha)$-quantile of $\Hmax(u_0)$. From this, (5.7) immediately follows. The statement (5.8) can be proven by the same arguments as for (A.13) in the proof of Theorem 5.3. \qed
\vspace{10pt}

\textbf{Proof of Corollary 5.5.} Let $q_{\alpha}(u_n)$ be the $(1-\alpha)$-quantile of $\Hmax(u_n)$ and $q_{\alpha}(u)$ the corresponding quantile of $\Hmax(u)$. We first show that for any $\alpha > 0$,
\begin{equation}\label{conv-quantiles-1} 
q_{\alpha}(u_n) \rightarrow q_{\alpha}(u)
\end{equation}
as $u_n \rightarrow u$. To do so, let $\mathcal{C}_u(\Delta,d)$ denote the space of uniformly continuous functions on $(\Delta,d)$ and define the functionals
\begin{align*}
 & M_n(x) = M_{u_n}(x) = \sup_{f \in \mathcal{F}} \sup_{0 \le w \le v \le u_n} | x(v,w,f)| \\
 & M(x) = M_{u}(x) = \sup_{f \in \mathcal{F}} \sup_{0 \le w \le v \le u} | x(v,w,f)| 
\end{align*}
for $x \in \mathcal{C}_u(\Delta,d)$. Elementary arguments show that 
\[ M(x) = \lim_{n \rightarrow \infty, y \rightarrow x} M_n(y), \]
where $x$ and $y$ are elements of $\mathcal{C}_u(\Delta,d)$. Using this together with the extended continuous mapping theorem (see e.g.\ Theorem 1.11.1 in \cite{vanderVaart1996}), we obtain that 
\[ M_n(H) \convd M(H). \]
Noting that $M_n(H) = \Hmax(u_n)$ and $M(H) = \Hmax(u)$, this can be re-expresses as
\[ \Hmax(u_n) \convd \Hmax(u).  \]
As the distribution function of $\Hmax(u)$ is continuous on $[0,\infty)$ by \ref{fact1}, we can conclude that the quantile functions converge as well, thus arriving at \eqref{conv-quantiles-1}.

Next let $\tilde{u}_0$ be a consistent estimator of $u_0$. By \eqref{conv-quantiles-1}, the quantile function $q_{\alpha}(\cdot)$ is continuous at each point $u$, in particular at $u_0$. Hence, 
\begin{equation}\label{conv-quantiles-2}
\hat{\tau}_{\alpha} = q_{\alpha}(\tilde{u}_0) \convp \tau_{\alpha} = q_{\alpha}(u_0). 
\end{equation}
Moreover, 
\begin{align*}
\pr \bigl( \hat{u}_0(\hat{\tau}_{\alpha}) < u_0 \bigr)
 & \le \pr \bigl( \sqrt{T} \hat{\Dsup}_T(u) > \hat{\tau}_{\alpha} \text{ for some } u < u_0 \bigr) \\
 & \le \pr \bigl( \sqrt{T} \hat{\Dmax}_T(u_0) > \hat{\tau}_{\alpha} \bigr) \\
 & = \pr \bigl( \hat{\Hmax}_T(u_0) > \hat{\tau}_{\alpha} \bigr). 
\end{align*}
Since $\hat{\Hmax}_T(u_0) \convd \Hmax(u_0)$ and the distribution function of $\Hmax(u_0)$ is continuous on $[0,\infty)$ by \ref{fact1}, the distribution function of $\hat{\Hmax}_T(u)$ uniformly converges to that of $\Hmax(u)$ on $[0,\infty)$. Hence,
\begin{align*}
\pr \bigl( \hat{\Hmax}_T(u_0) > \hat{\tau}_{\alpha} \bigr) 
 & = \pr \bigl( \Hmax(u_0) > \hat{\tau}_{\alpha} \bigr) + \Big[ \pr \bigl( \hat{\Hmax}_T(u_0) > \hat{\tau}_{\alpha} \bigr) - \pr \bigl( \Hmax(u_0) > \hat{\tau}_{\alpha} \bigr) \Big] \\
 & = \pr \bigl( \Hmax(u_0) > \hat{\tau}_{\alpha} \bigr) + o_p(1).
\end{align*}
Finally, as $\hat{\tau}_{\alpha} = \tau_{\alpha} + o_p(1)$ and the distribution function of $\Hmax(u_0)$ is continuous by \ref{fact1}, we obtain that 
\[ \pr \bigl( \Hmax(u_0) > \hat{\tau}_{\alpha} \bigr) = \pr \bigl( \Hmax(u_0) > \tau_{\alpha} \bigr) + o(1) = \alpha + o(1). \]
This completes the proof of (5.9). The statement (5.10) can again be shown by the same arguments as for (A.13) in the proof of Theorem 5.3. \qed

\setlength{\bibsep}{1pt}
\begin{small}
\bibliographystyle{ims}
\bibliography{bibliography}
\end{small}